**Title: Gigantic current control of coercive field and magnetic memory based on nm-thin ferromagnetic van der Waals $Fe_3GeTe_2$**

*Kaixuan Zhang*, Seungyun Han, Youjin Lee, Matthew J. Coak, Junghyun Kim, Inho Hwang, Suhan Son, Jeacheol Shin, Mijin Lim, Daegeun Jo, Kyoo Kim, Dohun Kim, Hyun-Woo Lee*, and Je-Geun Park**

Dr. Kaixuan Zhang, Youjin Lee, Dr. Matthew J. Coak, Junghyun Kim, Inho Hwang, Suhan Son, Prof. Je-Geun Park
Center for Correlated Electron Systems, Institute for Basic Science, Seoul 08826, South Korea
E-mail: zaal@mail.ustc.edu.cn; jgpark10@snu.ac.kr

Dr. Kaixuan Zhang, Youjin Lee, Dr. Matthew J. Coak, Junghyun Kim, Inho Hwang, Suhan Son, Jeacheol Shin, Prof. Dohun Kim, Prof. Je-Geun Park
Department of Physics and Astronomy, and Institute of Applied Physics, Seoul National University, Seoul 08826, South Korea

Dr. Kaixuan Zhang, Youjin Lee, Junghyun Kim, Inho Hwang, Suhan Son, Prof. Je-Geun Park
Center for Quantum Materials, Seoul National University, Seoul 08826, South Korea

Seungyun Han, Mijin Lim, Daegeun Jo, Prof. Hyun-Woo Lee
Department of Physics, Pohang University of Science and Technology, Pohang 37673, South Korea
E-mail: hwl@postech.ac.kr

Dr. Kyoo Kim
Korea Atomic Energy Research Institute, 111 Daedeok-daero, Daaejeon 34057, South Korea

Prof. Hyun-Woo Lee
Asia Pacific Center for Theoretical Physics, 77 Cheongam-ro, Nam-gu, Pohang 3773, South Korea

**Abstract:** Controlling magnetic states by a small current is essential for the next-generation of energy-efficient spintronic devices. However, it invariably requires considerable energy to change a magnetic ground state of intrinsically quantum nature governed by fundamental Hamiltonian, once stabilized below a phase transition temperature. We report that surprisingly an in-plane current can tune the magnetic state of nm-thin van der Waals ferromagnet $Fe_3GeTe_2$ from a hard magnetic state to a soft magnetic state. It is the direct demonstration of the current-induced substantial reduction of the coercive field. This surprising finding is possible because the in-plane current produces a highly unusual type of gigantic spin-orbit torque for $Fe_3GeTe_2$. And we further demonstrate a working model of a new nonvolatile magnetic memory based on the principle of our discovery in $Fe_3GeTe_2$, controlled by a tiny current. Our findings open up a new window of exciting opportunities for magnetic van der Waals materials with potentially huge impacts on the future development of spintronic and magnetic memory.

**Main Text:** The last few years have witnessed that the new arrival of two-dimensional (2D) magnetic van der Waals (vdW) materials have attracted colossal attention worldwide[1-6]. When combined with a wide variety of other 2D nonmagnetic vdW materials[7], these magnetic vdW materials can open up a new horizon of novel nano-electronic devices[8-11] that consist entirely of 2D materials. Among all the magnetic vdW materials, $Fe_3GeTe_2$ (FGT) received special attention because it is the only topological ferromagnetic vdW metal[12]. The nanoscale FGT is a metallic hard ferromagnet with a large coercivity of ~ several kOe[13]. It exhibits a significant anomalous Hall effect (AHE)[12,14], allowing easy probing of its magnetic configurations via electrical transport measurements. Besides, exciting new behavior was also found in the FGT-based heterostructure, such as the unconventional 3-state magnetoresistance [15].

Electrical modulation of the magnetic anisotropy or coercivity is critical to realizing energy-efficient spintronic devices such as memory. For information storage, the magnetic anisotropy should be high enough to minimize the information loss by fluctuations. In contrast, the magnetic anisotropy should be small for information writing with low energy consumption, which is desirable for any energy-efficient device. These self-contradicting conditions require that the anisotropy should be able to be lowered only on demand (current applied for writing), and otherwise kept high enough to ensure the stability of the information stored (no current for storage). Here we provide the first experimental demonstration for such on-demand modulation of the coercivity and a working model of a magnetic memory in nm-thin FGT.

In this experiment, we discover that the coercive field $H_c$ of FGT can be reduced by ~50% (~100%) by applying a tiny in-plane electric field of ~0.1 mV/nm (~0.3 mV/nm), respectively. In comparison, all the previous attempts used an out-of-plane electric field (often in the range of 0.1 V/nm), several orders of magnitude larger than ours, to induce a much smaller degree of modulation (e.g., <5% for FePt and FePd[16]): for example, semiconducting vdW magnets[17-20] and metallic magnetic thin films[16,21]. The coercivity modulation by an in-plane field has a drawback of energy dissipation. Nevertheless, our result of the monotonic modulation of the coercivity provides a promising alternative for electrical modulation. It requires neither a complicated ionic-liquid gating[14] nor a very high gating voltage[17-21] (generally > 30 V). Moreover, while the out-of-plane field case depends on a field-induced change of carrier density[16-20] or orbital occupation[21], we find that the electrical modulation of $H_c$ observed in FGT is strongly influenced by the spin-orbit torque (SOT)[22-24] induced by an in-plane electric current. The SOT is a current-generated torque acting on a magnetic order and usually generated by an in-plane current flowing in heavy metal (such as Pt and Ta) in contact with magnets[22,25,26]. In FGT, on the other hand, an in-plane current in FGT itself generates the SOT[27] owing to the particular geometrical structure of FGT. We estimate that the effective magnetic field due to the SOT is on the order of 50 Oe for a small current density of 1 mA/μm$^2$, which is about two orders of magnitude larger than that produced by heavy metals like Pt and Ta[28]. We further reveal that the large SOT is due to the large Berry curvature in the topological ferromagnetic metal FGT. Based on these observations, a prototypical nonvolatile magnetic memory of high energy-efficiency is further demonstrated in this work.

*Sample preparation and basic transport properties:* FGT single crystals were grown by the chemical vapor transport method and adopted for the device fabrication (see the Experimental section). Our bulk FGTs are hole-doped $Fe_xGeTe_2$ with $x$ varying from 2.77 to 2.89 according to the energy-dispersive X-ray spectroscopy (EDS) measurements. We made several FGT devices with different thicknesses (samples S1, S2, S3, S4, S5, and S6 with a thickness of 21.3, 16.7, 6, 42, 17.5, and 15 nm, respectively). **Figure 1**a shows a typical FGT device (sample S1) with electrodes of a Hall bar geometry. The thickness is measured to be 21.3 nm by the atomic force microscopy (AFM) in Figure 1b. Figure 1c displays the longitudinal resistance ($R_{xx}$) of FGT as a function of temperature $T$. The $R_{xx}$-$T$ curve exhibits the spin-flip scattering-induced magnetic transition with $T_c$ ~185 K and the Kondo-like minimum at ~20 K, both of which are the typical features of FGT[13]. Figure 1d depicts the transverse Hall resistance ($R_{xy}$) of FGT as a function of an out-of-plane magnetic field ($H$). The almost rectangle-shaped sharp hysteresis loop in the $R_{xy}$-$H$ curve confirms that FGT is indeed a hard magnet: the dominant contribution to $R_{xy}$ comes from the AHE[12,14] while the ordinary Hall effect is negligible. The $H_c$ of all our FGT nanoflakes varies from 1 to 2.2 kOe, which is indicative of hole-doped $Fe_xGeTe_2$ with $2.7<x<3$ according to previous investigations[13,29], and thus consistent with the EDS results.

*Current-dependent magnetism:* Interestingly, $H_c$ of the hysteresis loop gets significantly reduced as an in-plane current $I$ applied to FGT increases from 0.05 to 1.5 mA (**Figure 2**a). On the other hand, the remnant Hall resistance $R_{xy}^r$ remains almost unchanged with $I$, implying that the saturation magnetization of FGT is not much affected by $I$. Thus, $I$ reduces $H_c$ without disrupting the ferromagnetic ordering of FGT, which is ideal for spintronic device applications. $H_c$ and $R_{xy}^r$ of the sample S1 are shown in Figure 2b as a function of $I$. Note that the current of 1.5 mA reduces $H_c$ by more than 50% (cf. <5% for FePt and FePd[16]). We also measured several other samples (samples S2, S3, S4, and S5) with a similar reduction of $H_c$ by $I$ (Figure S1 for sample S2, Figure S2 and Figure S3 for sample S3, Figure S4 for sample S4, and Figure S5 for sample S5). The current dependence of $H_c$ for all the samples is summarized in Figure 2c. Figure 2d displays the normalized $H_c$ as a function of current density $J$, where $H_c$ is normalized by its value at small $J$ (~1 mA/μm$^2$). As can be seen, the current reduces $H_c$ by ~50% for ~7 mA/μm$^2$ (corresponding to ~0.07 mV/nm) and ~100% for ~30 mA/$\mu$m$^2$ (corresponding to ~0.3 mV/nm).

*Spin-orbit torque effect:* The $H_c$ reduction by the current can arise partly by the Joule heating since $H_c$ gets reduced at higher temperatures. We adopt three different methods to quantify the Joule heating contribution to the $H_c$ reduction. After comprehensive and systematic studies as summarized in the Supporting Information, we conclude that the Joule heating is responsible only for 30% (thickest sample S4) to 60% (thinnest sample S3) of the total $H_c$ reduction at $J$~10 mA/$\mu$m$^2$. It implies that roughly half of the reduction experimentally observed is due to a non-Joule-heating effect of nontrivial origin.

We would now like to examine how the non-Joule-heating effect can arise from the SOT. We demonstrate below the effective field due to the SOT in FGT is about two orders of magnitude larger than those values reported before for heavy metals such as Pt[28]. An in-plane current density $\mathbf{J}=(J_x, J_y, 0)$ generates the SOT $\mathbf{T}_{SOT}=-|\gamma|\mathbf{M}\times\mathbf{H}_{SOT}$ acting on the magnetization $\mathbf{M}=(M_x, M_y, M_z)$, where $|\gamma|$ is the gyromagnetic ratio. The symmetries of a given system constrain the structure of the effective field $\mathbf{H}_{SOT}$ in $\mathbf{T}_{SOT}$. **Figure 3**a shows the crystal structure of a monolayer $Fe_3GeTe_2$ with the distinct three symmetries: the three-fold rotation symmetry around the c-axis ($z$-axis), the mirror reflection symmetry with respect to the a-axis ($y$-axis), and the mirror reflection symmetry with respect to the c-axis. Under this set of symmetries, the general form of $\mathbf{H}_{\mathrm{SOT}}$ can be expressed as[27];

$$\mathbf{H}_{\mathrm{SOT}} = \Gamma_0 \left[ \left( m_x J_x - m_y J_y \right) \hat{\mathbf{x}} - \left( m_y J_x + m_x J_y \right) \hat{\mathbf{y}} \right],$$

where $\mathbf{m}=(m_x, m_y, m_z)= \mathbf{M}/|\mathbf{M}|$ and $\Gamma_0$ is a coefficient that parametrizes a magnetoelectric coupling strength. Unlike $\mathbf{H}_{\mathrm{SOT}}$ in other systems, the effective field $\mathbf{H}_{\mathrm{SOT}}$ of FGT is conservative[27] and can be obtained from an effective free energy density $f_{\mathrm{SOT}}$ by $\mathbf{H}_{\mathrm{SOT}} = -\partial f_{\mathrm{SOT}} / \partial \mathbf{M}$ with

$$f_{\mathrm{SOT}} = M_{\mathrm{s}} \Gamma_0 \left[ J_y m_x m_y - \frac{1}{2} J_x \left( m_x^2 - m_y^2 \right) \right].$$

It implies that the SOT effect in FGT amounts to the magnetic anisotropy change, whereas such an interpretation is not possible for the damping-like SOT in other systems. This structure of $f_{\mathrm{SOT}}$ derived for a monolayer FGT is also applicable to each layer of a multilayer FGT like our samples since each layer satisfies the symmetries, and the interlayer coupling is weak in the vdW FGT[30,31] (see more details in the Supporting Note 3). Together with the inherent free energy density $f_0 = -(1/2)K_z M_z^2 / M_s$ that describes the Ising-type perpendicular magnetic anisotropy of FGT, one obtains the effective free energy density $f_{\mathrm{eff}} = f_0 + f_{\mathrm{SOT}}$ in the presence of current as follows:

$$f_{\mathrm{eff}} = -\frac{M_{\mathrm{s}}}{2} \left[ K_z \cos^2 \theta + \Gamma_0 J \sin^2 \theta \cos \left( 2\phi + \phi_J \right) \right],$$

where the angles $\theta$ and $\phi$ specify the directions of the unit vector $\mathbf{m}=(\sin\theta \cos\phi, \sin\theta \sin\phi, \cos\theta)$, and $\mathbf{J}=J(\cos\phi_J, \sin\phi_J, 0)$.

We now demonstrate how $J$ can effectively reduce the barrier height between the local minima of the free energy. Figure 3b shows our theoretical calculations for the effective free energy density $f_{\rm eff}$ profile when the current is applied along the armchair direction ($\phi_J$=0). For $J$=0, $f_{\rm eff}$ has a minimum value of $-M_sK_z/2$ at $\mathbf{m}=\pm\hat{\mathbf{z}}$ and a maximum value of 0 when $\mathbf{m}$ lies in the $xy$ plane (left panel). Thus to switch the magnetization from $\mathbf{m}=-\hat{\mathbf{z}}$ to $\mathbf{m}=+\hat{\mathbf{z}}$, one should overcome the free energy barrier of $M_sK_z/2$. Then the $f_{\rm eff}$ profile for $\Gamma_0 J > 0$ (middle panel) shows that interestingly the free energy barrier along $\mathbf{m}=\pm\hat{\mathbf{x}}$ is decreased. Although the barrier is increased along $\mathbf{m}=\pm\hat{\mathbf{y}}$, this increase is irrelevant since a magnetic switching occurs through a path that minimizes the energy barrier, which in this case is $\mathbf{m}= -\hat{\mathbf{z}}\rightarrow \pm\hat{\mathbf{x}} \rightarrow +\hat{\mathbf{z}}$. Likewise, the $f_{\rm eff}$ profile for $\Gamma_0 J < 0$ (right panel) exhibits that the free energy barrier is lowered along $\mathbf{m}=\pm\hat{\mathbf{y}}$, resulting in the switching along the path $\mathbf{m}=-\hat{\mathbf{z}} \rightarrow \pm\hat{\mathbf{y}} \rightarrow +\hat{\mathbf{z}}$. A simple analysis shows that $J$ modifies the free energy barrier from $M_sK_z/2$ to $M_sK_z/2 - M_s\left|\Gamma_0 J\right|/2$ (see the Experimental Section). The free energy barrier can be lowered further by $H$ and vanishes when $H = K_z - \left|\Gamma_0 J\right|$. Thus $H_c$ is given by $K_z - \left|\Gamma_0 J\right|$ for a single domain case. This analysis shows that $H$ and $\left|\Gamma_0 J\right|$ play similar roles in the energy barrier. In an actual switching process, however, the switching can be more complicated due to the magnetic domains[32] in FGT, and the free energy barrier for $H$=$J$=0 becomes smaller[33,34] than $K_z$. Although it is difficult to have an analytic evaluation of the free energy barrier, we can approximate it by the experimentally measured $H_c(J=0)$ since $H$ competes with the free energy barrier and makes the switching when $H$=$H_c(J=0)$[35]. Considering that $H$ and $|\Gamma_0 J|$ play similar roles for the energy barrier, $H_c(J)$ can be approximated by $H_c(J)=H_c(0) - |\Gamma_0 J|$ and we then obtain the theoretical prediction $H_c(0) - H_c(J)=\Delta H_c(J)=|\Gamma_0 J|$ (solid blue line in Figure 3c), which was also adopted in the previous studies[35].

The predicted linear dependence on $|J|$ is roughly in agreement with the linear-in-$J$ behavior of $\Delta H_c(J)$ (Figure 3d) in the relatively small $J$ regime. By fitting the experimental result with the above formula, one obtains the first estimation of the spin-orbit field $|\mathbf{H}_{\mathrm{SOT}}|\sim|\Gamma_0 J|$, ranging from 75 to 135 Oe for $J$=1 mA/μm$^2$ (Figure 3e). It is an overestimation, however, since the $H_c$ reduction is also partly due to the Joule heating. After carefully excluding the Joule-heating contribution, we obtain the refined and more realistic estimate of the non-Joule-heating contribution $|\mathbf{H}_{\mathrm{SOT}}|\sim|\Gamma_0 J|$, which is 50±15 Oe for $J$=1 mA/μm$^2$ for all the samples (Figure 3e). Please note that this effective field strength (per $J$) is two orders of magnitude larger than that produced by heavy metals such as Pt and Ta[28], and comparable only to that produced by the topological insulator BiSb[35] (Figure 3f). We also estimate $|\mathbf{H}_{\mathrm{SOT}}|$ independently by measuring the anisotropic magnetoresistance (AMR) as a function of an in-plane magnetic field direction. Since SOT makes the effective free energy dependent on the azimuthal angle of the magnetization (Figure 3b), the AMR can deviate from its conventional angular dependence. This predicted deviation is observed in our experiments, and by fitting the difference, one obtains $|\Gamma_0 J|\sim119$ Oe for $J$=1 mA/$\mu$m$^2$ (Supporting Note 2). To get further confidence in our analysis, we performed the first-principles calculation to evaluate the Berry curvature contribution to $\mathbf{H}_{\mathrm{SOT}}$ (Supporting Note 3) and obtain $|\Gamma_0 J|\sim30$ Oe at $J$=1 mA/$\mu$m$^2$ for hole-doped $Fe_xGeTe_2$ with $2.7<x<3$. From these three independent analyses that produce similar estimations of $\Gamma_0$, we conclude that the non-Joule-heating-effect contribution to the current-induced $H_c$ reduction arises from the SOT. It was reported[12] that FGT is a topological nodal-line semimetal with the topological band structure enhancing the Berry curvature contribution to the anomalous Hall effect. The considerable $\mathbf{H}_{\mathrm{SOT}}$ value estimated from our analysis can share a similar origin. By the further comprehensive theoretical calculations and then careful comparison with the experimental results, we concluded that most of our observed SOT effects ought to be intrinsic (see more discussions in Supporting Note 3). It is not easy to explicitly evaluate whether there is some remaining extrinsic disorder contribution[36] at present, which might be clarified in further study.

A recent experiment[25] on the FGT/Pt bilayer system measured the SOT magnitude of the bilayer. The result is one order larger than that produced by the conventional heavy metal Pt and one order smaller than the SOT in our present work (Figure 3f). We suspect that for the bilayer, both Pt and FGT may affect the measured SOT magnitude. Considering the current shunt in Pt, the intermediate SOT magnitude of FGT/Pt system between that produced by Pt and that produced by FGT provides a shred of consistent evidence for such conjecture.

When $J$ becomes larger and closer to $J_c$ ($\equiv H_c(0)/|\Gamma_0|$), i.e., the effective energy becomes small, multi-domain structure effect also gets severe during the magnetization switching. In such a situation, the hysteresis loop may be no longer as sharp as the case with a small current, consistent with the experimental results (Figure 2a and Figure S3). In systems with the perpendicular magnetic anisotropy, magnetic dipolar interaction between neighboring domains can effectively reduce the magnetic field acting on the system, thereby enhancing $H_c$ for bigger current. The red circles in Figure 3c depict the enhanced $H_c$, which is similar to the experimental result (Figure 3d).

Note that for conventional ferromagnetic materials like FGT, the magnetic anisotropy decreases as temperature increases, and thus the coercivity generally reduces upon increasing temperature[13]. Joule heating can enhance the temperature of the device and, therefore, can correspondingly reduce the coercivity without requiring spin-orbit torques. However, the Joule heating effect frequently exists in many SOT systems[25] associated with large current. It is not suitable for device performance considering the stability of working temperature and the device aging problem, etc., although it can reduce the coercivity. Therefore, it is always valuable and desirable to find some new systems like FGT with large spin-orbit torques. On the other hand, for better practical application, we need to decrease or control the Joule heating effect of the device in the future.

*Robust nonvolatile magnetic memory:* Next, we demonstrate a new type of magnetic memory based on our observation of the giant current-controlled coercivity reduction in nm-thin vdW FGT. **Figure 4**a shows the schematic of the magnetic memory device, where the information can be easily written by writing the current $I_{\mathrm{write}}$ and read through $R_{xy}$. For instance, when the device is initially in the "0" state, we can alter it to the "1" state through the writing path I: we first turn on a small $H$ in the order of 100 Oe (e.g., 850 Oe) then apply a $I_{\mathrm{write}}$ (Figure 4b). Since $H_{\mathrm{c}}$ is significantly reduced by $I_{\mathrm{write}}$, the magnetization of the FGT device can be switched by current from "0" to "1" states above certain critical current.

Figure 4c shows the measured Hall voltage ($V_{xy}$) as a function of $I_{\mathrm{write}}$ (0→ 2→ -2 → 2 mA) under various $H$ from -0.5 to 0.5 T. From these $V_{xy}(H)$-$I_{\mathrm{write}}$ curves, we define $R_{xy}(H)$-$R_{xy}$ (-0.5 T) as ($V_{xy}(H)$-$V_{xy}$ (-0.5 T))/$I_{\mathrm{write}}$, and the converted results are shown in Figure 4d. It is clear that initially, the FGT device is robustly in the "0" state ("-$M_{\mathrm{s}}$" state) regardless of current sweep under -0.5, and 0 T; then under 0.085 T, when $I_{\mathrm{write}}$ increases from 0 to 2 mA, the magnetization gradually increases to "+$M_{\mathrm{s}}$" (indicated by the blue arrow), i.e., to the "1" state. It is noteworthy that after reaching the "1" state, the magnetization is well maintained while sweeping the current. It proves that such magnetic information transition from "0" to "1" states through $I_{\mathrm{write}}$ is robust and nonvolatile.

Although still prototypical, we can emphasize several significant merits of the FGT-based device. First and foremost, it is a first working memory device entirely based on the vdW topological ferromagnetic metal (FGT) with the unusual type of gigantic SOT. Second, thanks to the matured technique for 2D materials, FGT-based devices can, in principle, be fabricated without undesired defects, including many grain boundaries and point defects, which are easily introduced during the growth and annealing processes for the thin film devices like FeCoB/MgO/FeCoB[37,38]. More importantly, such an FGT-based magnetic memory device is highly energy-efficient. For example, the well-known FeCoB/MgO/FeCoB based SOT-MRAM[39] requires an extremely high current density of >400 mA/μm$^2$ to switch the magnetization, which is about 80 times larger than that in our FGT device of ~5 mA/μm$^2$ (corresponds to ~1 mA here).

In summary, we demonstrate the current control of the coercive field for nm-thin FGT, a 2D vdW topological ferromagnetic metal. We show that the observed current control is consistent with the unique properties of the SOT in FGT, and the gigantic SOT field is mainly due to the large Berry curvature of this topological material. We also successfully present a novel type of robust nonvolatile magnetic memory based on the giant SOT effect of FGT by using a tiny current. Our findings open up a fascinating avenue of electrical modulation and spintronic applications using 2D magnetic vdW materials.

**Experimental Section**

*Sample growth and device fabrication:* FGT single crystals were grown by the chemical vapor transport method with iodine as the transport agent. High-purity elements (Fe, Ge, Te) were stoichiometrically mixed and sealed in an evacuated quartz tube. The crystals were grown in a two-zone furnace with a temperature gradient of 750 ˚C (source) to 680 ˚C (sink) for 7 days.

To minimize the sample oxidation and degradation, all our FGT nanoflakes were exfoliated from the as-grown single crystals onto 285 nm-$SiO_2$/Si substrates inside the glove box ($O_2$: <0.6 ppm; $H_2O$: <0.2 ppm) by a conventional mechanical exfoliation method with Scotch tape. Before taken out of the glove box, the samples were sealed in a vacuum plastic package to prevent degradation due to air. Before the standard electron beam lithography (EBL), the spin-coated PMMA polymer on FGT was prebaked with a relatively moderate condition: 130 ˚C for 1.5 min.

To reduce the sample oxidation and degradation of FGT nanoflake, 90/5 nm Au/Ti electrodes were deposited by the electron beam evaporation under a high vacuum (<$10^{-5}$ Pa) immediately after EBL. During the whole process, including the transport measurements, the FGT samples were exposed to air for less than ~25 min. Under this condition, the successfully measured FGT samples displayed typical and prominent $R_{xy}$-$H$ hysteresis loops, without visible signs of considerable damage. In contrast, when the FGT sample was exposed to air for over 70 min in one of our control experiments, the sample was severely damaged. The measured $R_{xx}$-$T$ curve shows a significantly insulating behavior, and the $R_{xy}$-$H$ loop is too noisy to be a well-defined ferromagnetic hysteresis loop (Figure S9).

We fabricated several FGT devices (samples S1, S2, S3, S4, S5, and S6 with a thickness of 21.3, 16.7, 6, 42, 17.5, and 15 nm, respectively). For the FGT/$NbSe_2$ heterostructure (sample S4), the $NbSe_2$ nanoflake was exfoliated onto a PDMS stamp and then dry transferred onto the $SiO_2$/Si substrate to form the FGT/$NbSe_2$ heterostructure.

*Electrical transport measurements:* Transport measurements with a Hall-bar geometry were made using a Quantum Design physical property measurement system with the highest magnetic field up to 9 T. Gold wires were used to make contacts between the chip carrier and the Au/Ti electrodes.

*Theoretical calculations:* This section presents the calculated free energy barrier $M_s K_z^{\text{eff}}/2$ for the magnetization switching from $\mathbf{m}=-\hat{\mathbf{z}}$ to $\mathbf{m}=+\hat{\mathbf{z}}$. From the effective free energy density

$$f_{\text{eff}} = -\frac{M_s}{2}\left[K_z \cos^2\theta + \Gamma_0 J \sin^2\theta \cos\left(2\phi + \phi_J\right)\right],$$

it is evident that $J$ does not affect $f_{\text{eff}}$ for $\mathbf{m}=+\hat{\mathbf{z}}$ ($\theta = 0$) and $\mathbf{m}=-\hat{\mathbf{z}}$ ($\theta = \pi$). On the other hand, $f_{\mathrm{eff}}$ depends on $\phi$ for $\theta \neq 0, \pi$ (Figure 3b); $f_{\mathrm{eff}}$ as a function of $\phi$ is minimized for $\cos(2\phi + \phi_J) = \operatorname{sgn}(\Gamma_0 J)$ and maximized for $\cos(2\phi + \phi_J) = -\operatorname{sgn}(\Gamma_0 J)$. Thus for the unique value of $\phi$ that minimizes $f_{\mathrm{eff}}$, $f_{\mathrm{eff}}$ becomes a function of $\theta$ only and is given by

$$f_{\mathrm{eff}} = -\frac{M_{\mathrm{s}}}{2}\left[K_z \cos^2\theta + \left|\Gamma_0 J\right| \sin^2\theta\right] = -\frac{M_{\mathrm{s}}}{2}\left[\left(K_z - \left|\Gamma_0 J\right|\right)\cos^2\theta + \left|\Gamma_0 J\right|\right].$$

From this, one finds the free energy barrier of $M_{\mathrm{s}} K_z^{\mathrm{eff}} / 2$ with the effective magnetic anisotropy $K_z^{\mathrm{eff}}$ given by

$$K_z^{\mathrm{eff}} = K_z - \left|\Gamma_0 J\right|.$$

**Supporting Information**

Supporting Information is available from the Wiley Online Library or the author.

*Note added.* During the preparation of our work, we found the reported modulation of coercivity in $Fe_3GeTe_2/WTe_2$ heterostructure in a very recent work[40].

**Acknowledgments**

We thank Kihoon Lee, Haleem Kim, and Nahyun Lee for their supports and helpful discussions. We also thank Dongjin Lee and Heedeuk Shin for their help with finite-element numerical calculations. IBS-CCES was supported by the Institute for Basic Science (IBS) in Korea (Grant No. IBS-R009-G1), and CQM was supported by the Leading Researcher Program of the National Research Foundation of Korea (Grant No. 2020R1A3B2079375). The theoretical works at the POSTECH were funded by SSTF (Grant No. BA-1501-07). K.K. was supported by the National Research Foundation (NRF) Korea (Contract No. 016R1D1A1B02008461).

Received: ((will be filled in by the editorial staff))
Revised: ((will be filled in by the editorial staff))
Published online: ((will be filled in by the editorial staff))

**Figures**

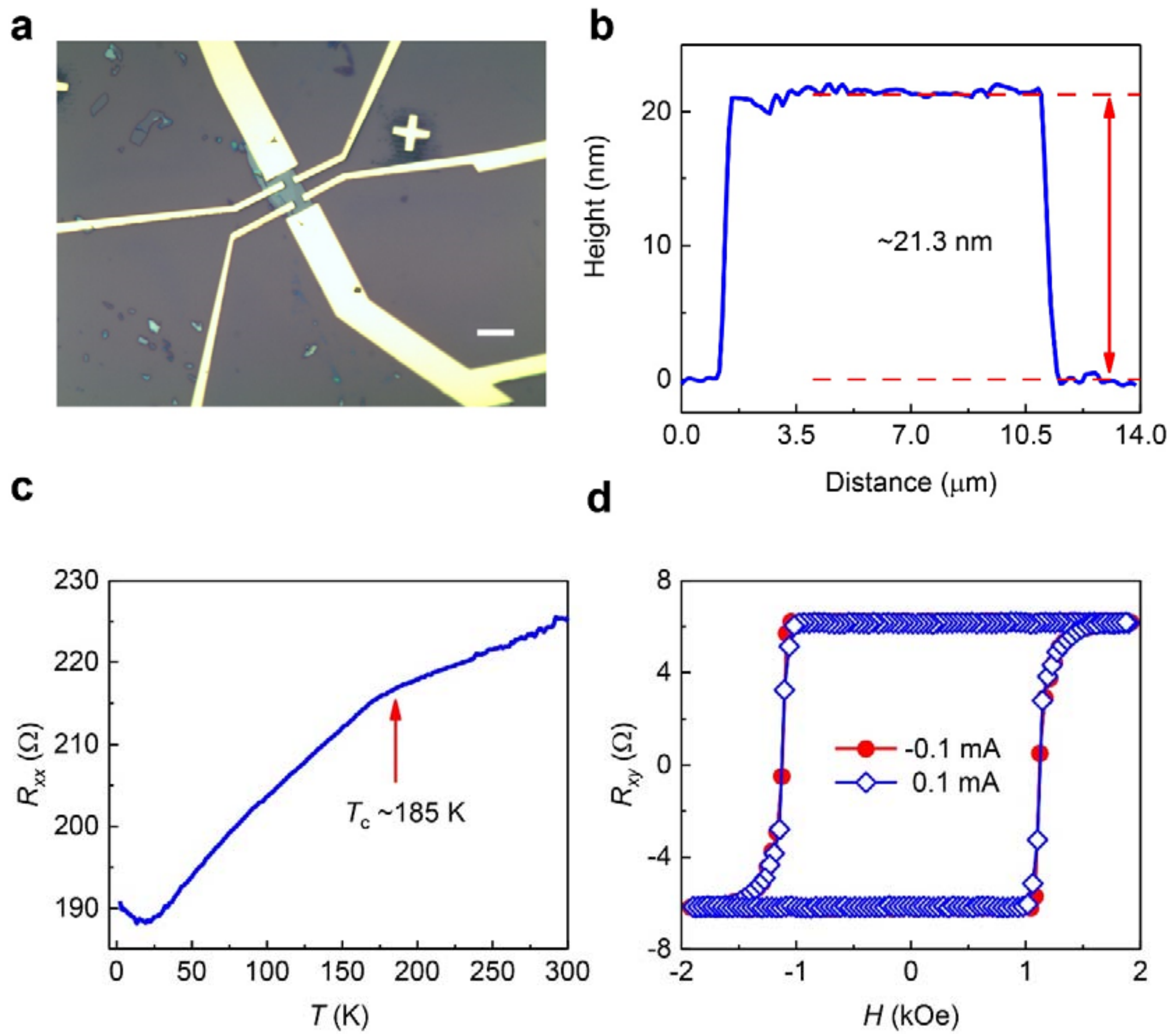

**Figure 1.** Sample preparation and transport measurements. a) Optical image of a typical FGT nanoflake sample (sample S1) with a Hall-bar geometry electrode. The white scale bar represents 10 μm. b) The thickness of sample S1 is 21.3 nm, as measured by AFM. c) Longitudinal resistance $R_{xx}$ as a function of temperature $T$ with current $I$=0.05 mA. The red arrow indicates the magnetic transition due to spin-flip scattering, from which $T_c$ ~ 185 K is determined. d) Hall resistance $R_{xy}$ as a function of magnetic field $H$ at 2 K with $I$= +0.1 mA and -0.1 mA.

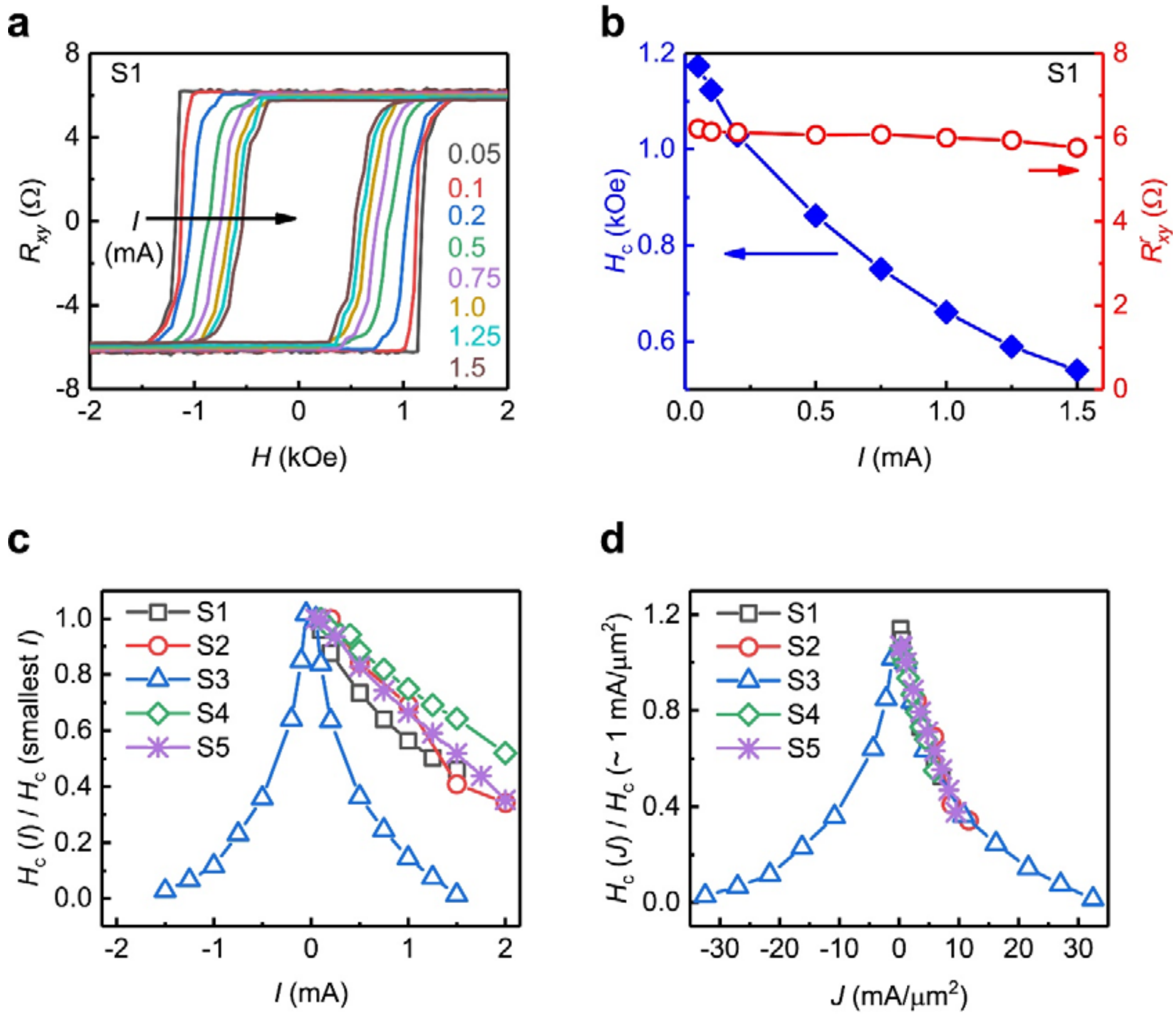

**Figure 2.** Current-dependent magnetism. a) $R_{xy}$-$H$ curves of sample S1 at 2 K with applied current $I$ varying from 0.05 to 1.5 mA. b) Extracted coercive field $H_c$ and remnant Hall resistance $R_{xy}{}^r$ of sample S1 as a function of applied current $I$ at 2 K. c) $H_c(I)/H_c$ (smallest $I$) as a function of applied current $I$ for sample S1 (21.3 nm; black square), S2 (16.7 nm; red circle), S3 (6 nm; blue triangle), S4 (42 nm; green diamond), and S5 (17.5 nm; purple star) at 2 K. d) $H_c$ ($J$) / $H_c$ (~1 mA/μm$^2$) as a function of current density $J$ for sample S1, S2, S3, S4, and S5 at 2 K.

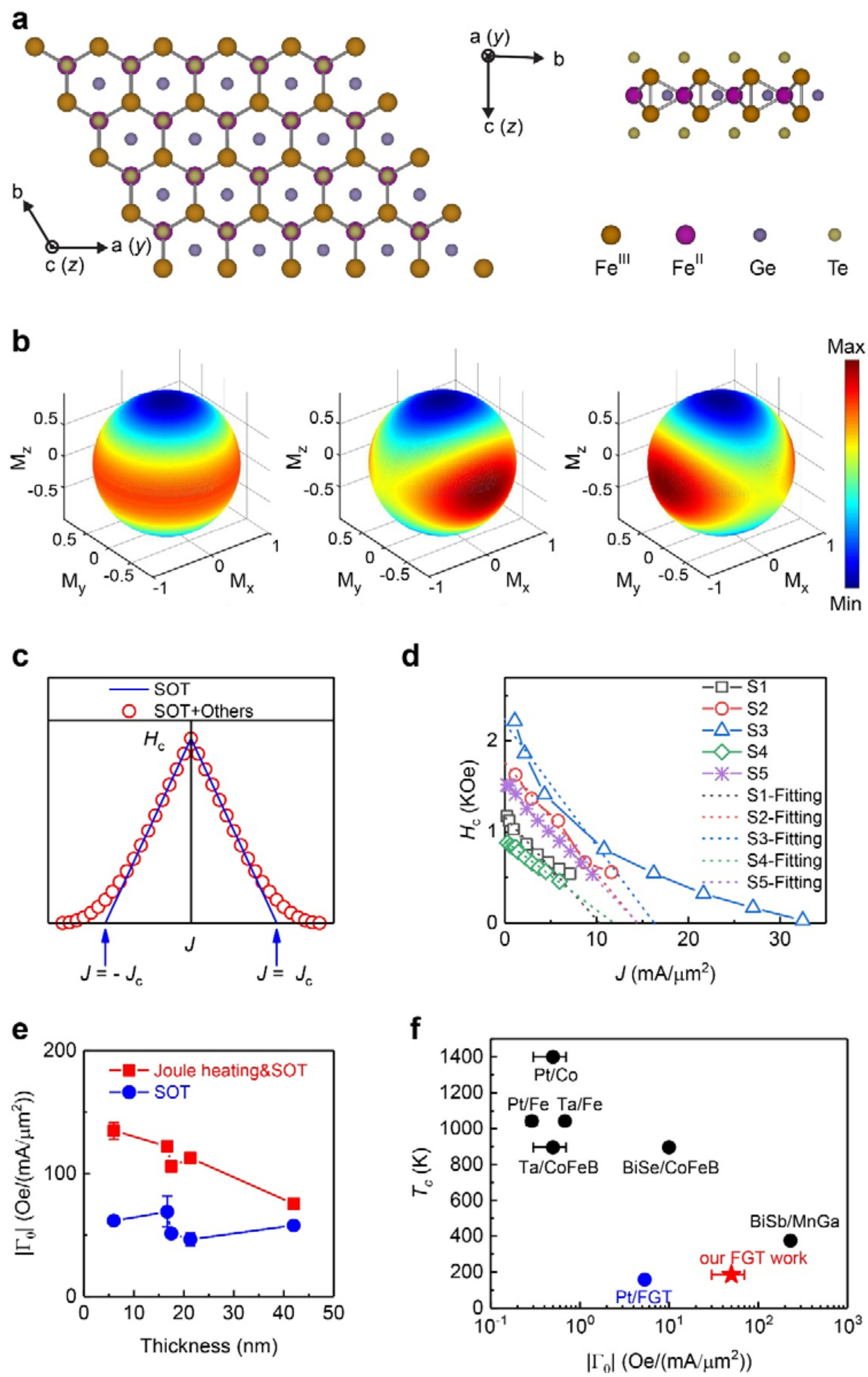

**Figure 3.** Analysis based on the spin-orbit torque of FGT. a) The crystal structure of a monolayer $Fe_3GeTe_2$. Left: View along the c axis; right: view along the a-axis. $Fe^{III}$ and $Fe^{II}$ denote the two

inequivalent Fe sites with oxidation number +3 and +2, respectively. b) Free energy profiles for $J$=0 (left panel), $J > 0$ (middle panel), and $J < 0$ (right panel). c) Theoretical curves for the coercivity $H_c$ as a function of the current density $J$. The solid blue lines indicate the results due to the SOT. In contrast, the open red circles indicate the results due to both SOT and other mechanisms, e.g., magnetic dipolar interaction between neighboring domains. d) Experimental $H_c$-$J$ curves at 2 K for sample S1 (black square), S2 (red circle), S3 (blue triangle), S4 (green diamond), and S5 (purple star). The dash lines denote the corresponding linear fitting lines within a relatively small $J$ range. e) Estimated $|\Gamma_0|$ for the SOT contribution alone (blue circles) and the combined case of both Joule-heating-effect and SOT contribution (red squares) as a function of the sample thickness. f) The values of $|\Gamma_0|$ and $T_c$ from our FGT samples and various composite SOT systems[25,28,35,41].

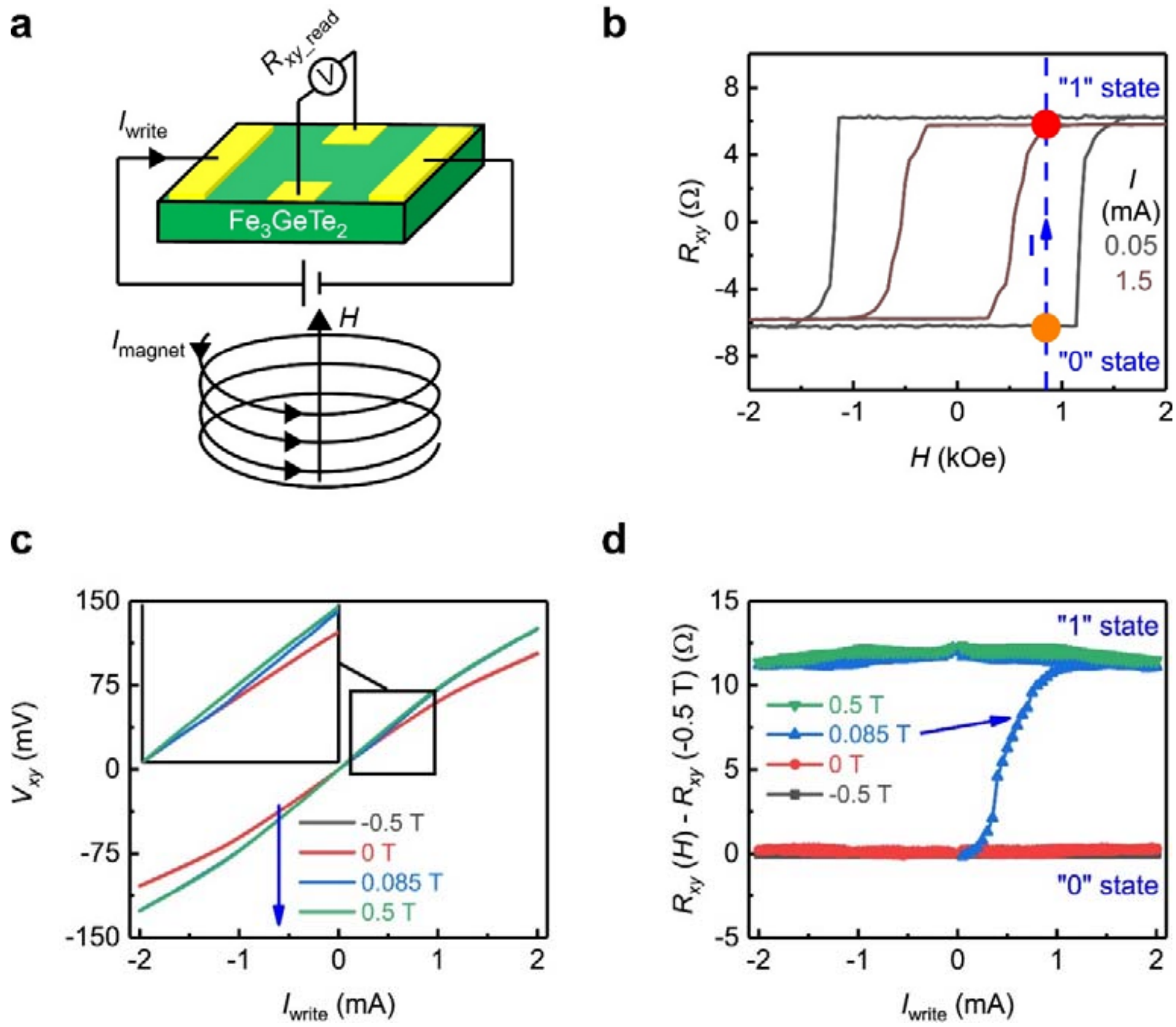

**Figure 4.** FGT-based magnetic memory. a) Schematic of the magnetic memory device. The magnetic information can be written by the current $I_{write}$ and read by transverse Hall resistance $R_{xy}$. b) For sample S1, it illustrates the transition from "0" to "1" state under 0.085 T. The blue arrows indicate the writing path I of magnetic information. c) Measured Hall voltage ($V_{xy}$) as a function of $I_{write}$ (0→ 2→ -2 → 2 mA) under various magnetic fields from -0.5 to 0.5 T (the blue arrow indicates the measurement sequence). The inset is a magnified view for the black box where the deviation under 0.085 T is clearer. d) $R_{xy}(H)$-$R_{xy}$ (-0.5 T) defined as ($V_{xy}(H)$-$V_{xy}$ (-0.5 T))/$I_{write}$ as a function of $I_{write}$ (0→ 2→ -2 → 2 mA) under various magnetic field from -0.5 T to 0.5 T. The blue arrow indicates the transition from "0" to "1" state while ramping $I_{write}$ from 0 to 2 mA under 0.085 T.

**The table of contents entry**

Substantial coercivity reduction by the current, larger at least by two orders of magnitude than those in previous reports, is found in van der Waals ferromagnet $Fe_3GeTe_2$. It is theoretically shown to arise from an unusual type of gigantic spin-orbit torque, which itself is directly related to its special symmetries, large Berry curvature, and band topology. We also produce a working model of a new robust nonvolatile magnetic memory based on $Fe_3GeTe_2$, controlled by a much smaller current. Our findings open up a new window of exciting opportunities for magnetic van der Waals materials with potentially huge impacts on the future spintronics.

Kaixuan Zhang*, Seungyun Han, Youjin Lee, Matthew J. Coak, Junghyun Kim, Inho Hwang, Suhan Son, Jeacheol Shin, Mijin Lim, Daegeun Jo, Kyoo Kim, Dohun Kim, Hyun-Woo Lee*, and Je-Geun Park*

Title: Gigantic current control of coercive field and magnetic memory based on nm-thin ferromagnetic van der Waals $Fe_3GeTe_2$

ToC figure

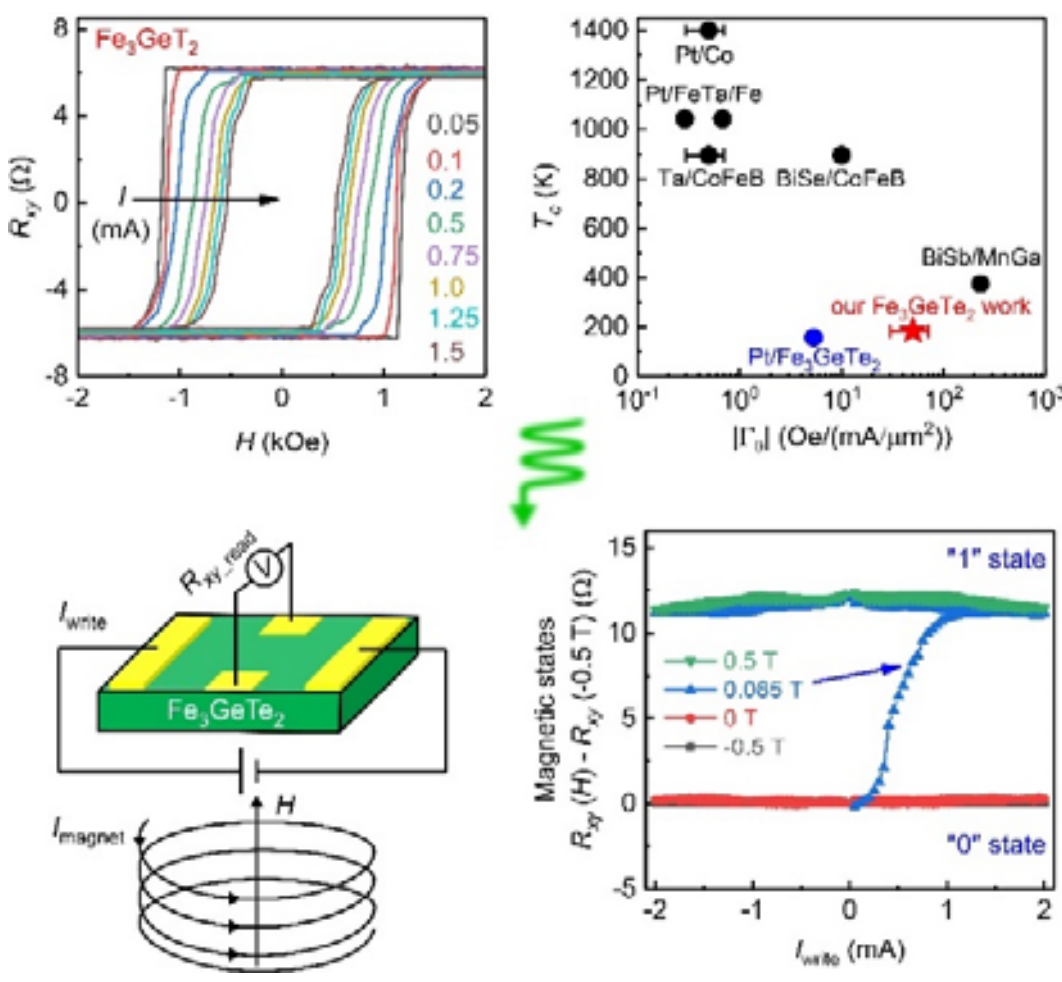

# Supporting Information

**Gigantic current control of coercive field and magnetic memory based on nm-thin ferromagnetic van der Waals $Fe_3GeTe_2$**

*Kaixuan Zhang*, Seungyun Han, Youjin Lee, Matthew J. Coak, Junghyun Kim, Inho Hwang, Suhan Son, Jeacheol Shin, Mijin Lim, Daegeun Jo, Kyoo Kim, Dohun Kim, Hyun-Woo Lee*, and Je-Geun Park**

## Contents:

## Supporting Notes:

## Supporting Figures:

**Supporting Notes**

1. **Three different methods to assess the Joule heating effect.**

The Joule heating can induce $H_c$ reduction, especially at a larger current. We thus have to evaluate its impact, which is not easy since it is impractical for us to use a commercial thermometer (e.g., Cernox-1070) to monitor the temperature of a nanofabricated FGT sample. It is difficult to very precisely evaluate Joule heating and get an accurate estimate of the temperature of a nanofabricated sample. Here we employ three different methods to find a coherent estimation of Joule heating and see whether there is a sizeable non-Joule-heating contribution to the observed $H_c$ reduction. The most direct approach uses the $R_{xx}$ of FGT as an internal thermometer, which is also the method adopted in a previous study[1] for the FGT/Pt system. For more confidence, we also took two other ways: one is using $NbSe_2$ as a nanofabricated thermometer and another the COMSOL simulations.

The schematic described in Figure S4a provides an experimental design to estimate the Joule heating effect by taking advantage of the high compatibility and design flexibility of 2D materials. For this, we made an FGT/$NbSe_2$ heterostructure (sample S4), where the red and blue dashed boxes indicate FGT and $NbSe_2$, respectively. The two ends of FGT and $NbSe_2$ were made physically attached to form a heterostructure, which was then covered by a big patchy of the electrode covering much of the overlapping region to ensure that FGT and $NbSe_2$ were in good thermal contact with each other. In this geometry, we can use $NbSe_2$ as a nanofabricated thermometer to monitor the FGT temperature. This method is hard to apply to 3D metal-based spin-orbit torque devices due to the lack of high flexibility in those samples, which is a unique advantage of our 2D magnetic vdW materials. This method can also be useful for future research on the current-dependent properties of other 2D magnetic materials.

Figure S4b shows the resistance of $NbSe_2$ as a function of temperature, $R(NbSe_2)$-$T$ curve, and Figure S4c shows the resistance of $NbSe_2$ while sweeping the current of FGT from 0 to 2 mA at 2 K in the $R(NbSe_2)$-$I$(FGT) curve. From Figure S4b,c, we obtained the calibrated $T$(FGT)-$I$(FGT) relationship at 2 K (Figure S4d). Note that the Joule heating induced a temperature change reaching ~24 K for the maximum current of 2 mA (corresponding to $J$~6 mA/$\mu$m$^2$) at 2 K: see

Figure S4e to compare the temperature- and current-induced reductions of $H_c$. As one can see, the $H_c$ reduction measured for $I$(FGT)=2 mA is comparable with the effect induced by temperature as high as 75 K, which is much higher than the temperature increase of 24 K estimated from the Joule heating effect. We thus conclude that the Joule heating cannot entirely be the primary origin of the current-induced $H_c$ reduction in this sample. Figure S4f shows how we separated the measured $H_c$ reduction into two contributions: one is the Joule-heating contribution and another the non-Joule-heating contribution of nontrivial origin.

We also performed a similar analysis using the $R_{xx}$ of FGT as an internal thermometer experimentally (sample S5; Figure S5), which is a more direct and straightforward method to assess the Joule heating (Supporting note 1.2). It also supports our conclusion that the Joule heating effect cannot fully explain the observed current-induced $H_c$ reduction.

As another independent check, we performed the finite-element numerical calculation using the COMSOL package. According to our calculations (Figure S6), the theoretically estimated temperature increases agree with the experimentally assessed ones for the samples S4 and S5 (Supporting note 1.3). Our calculations showed the following temperature increases for other samples with the maximal currents: 23.5 K for the sample S1 ($I$=1.5 mA with $J$~8 mA/$\mu$m$^2$), 43 K for the sample S2 (2.0 mA with $J$~11 mA/$\mu$m$^2$), and 62 K for the sample S3 (1.5 mA with $J$~32 mA/$\mu$m$^2$), respectively. By comparing the temperature- and current-induced reductions of $H_c$ for the samples S1 (Figure S7) and S3 (Figure S8) with these calculated temperature increases, we verify too that the Joule heating effect cannot fully explain the observed current-induced $H_c$ reduction (Supporting note 1.4). Our three different methods confirm a considerable coercivity reduction of nontrivial origin in addition to the simple Joule heating effects. For example, for the thinnest sample (sample S3) with the expected largest Joule heating effect at $J$~32 mA/$\mu$m$^2$, we found that the Joule heating effect is estimated to reduce $H_c$ from ~2220 to ~400 Oe while the non-Joule-heating effect is responsible for additional suppression of $H_c$ from ~400 to ~30 Oe. Note that the further reduction (from ~400 to ~30 Oe) estimated for the non-Joule-heating effect is still a very significant value compared to the modulation reported in FePt from ~1160 to ~1110 Oe by the ionic liquid gating[2].

We note that it is not ideal for predicting the effect of Joule heating by using FGT/ $NbSe_2$ heterostructure. And there can be a deviation between the COMSOL simulation and experimentally assessed temperature. Nevertheless, we are confident, based on our extensive experiments and calculations, that the differences of the estimated temperatures from the three different methods should be smaller than 10 K. Such deviation is much lower than the required temperature rises of ~60-125 K if Joule heating is the single origin for the $H_c$ reduction. Therefore, the specific inaccuracy of estimated temperatures by these two methods doesn't affect the conclusion that there is a sizeable non-Joule-heating contribution to the $H_c$ reduction.

Using these estimated temperatures, we can separate the $H_c$ reduction into the Joule-heating contribution and the SOT contribution. It then allows us to obtain the strength of the spin-orbit fields directly: with $J$=1 mA/$\mu$m$^2$, $|\mathbf{H}_{SOT}|\sim|\Gamma_0 J|\sim$105 Oe for the case of both Joule-heating and SOT contributions combined, and $|\mathbf{H}_{SOT}|\sim|\Gamma_0 J|\sim$50 Oe for the example of the SOT contribution alone (Figure 3e). It is noteworthy that this value of the spin-orbit field is huge, about two orders of magnitude larger than that produced by heavy metals such as Pt and Ta[3], and comparable to that produced by the topological insulator BiSb[4] (Figure 3f). Such estimates of $\Gamma_0$ are consistent with the following AMR results and the theoretical calculations based on the SOT model, which will be discussed later in the Supporting Information. When putting together, they collectively demonstrate the large SOT in FGT, which is one of the central claims of this work.

1.1 **FGT/$NbSe_2$ heterostructure: using $NbSe_2$ as a nanofabricated thermometer (sample S4).** As we discussed in detail above (Figure S4), we have developed an experimental design to quantify the Joule heating effect by taking advantage of the high compatibility and design flexibility of 2D materials. We fabricated an FGT/$NbSe_2$ heterostructure to use $NbSe_2$ as a nanofabricated thermometer for the sample temperature. This method is generally more challenging to achieve for conventional 3D metallic spin-orbit torque devices due to the lack of high flexibility in their samples, which fortunately is possible with our 2D magnetic van der Waals materials. We would also like to comment that this demonstration of FGT/$NbSe_2$ heterostructure will be useful for future research on current-dependent properties in other 2D magnetic materials. Utilizing this experiment, we conclude that there must be another effect at work with an enormous contribution, comparable to that of the Joule heating for the effect of current-induced $H_c$ reduction. Moreover, we can make a direct estimate of the nontrivial effect due to the net spin-orbit torque for the $H_c$ reduction after subtracting off the contribution by the Joule heating.

1.2 **Using the $R_{xx}$ of FGT as an internal thermometer (sample S5).**

One can use FGT itself as a thermometer. Since $R_{xy}$ is also affected by the SOT, we decided to use $R_{xx}$ as an internal thermometer, which is also the method adopted to assess the Joule heating effect of the FGT/Pt device in a very recent work[1]. Figure S5a,b shows the $R_{xx}$-$T$ and $R_{xx}$ –$I$ curves of the sample S5, from which the temperature of FGT is read or calibrated as a function of applied current. It can be seen that as the current increases to 2 mA, the temperature rises gradually to 40 K.

We then compared the temperature- and current-induced reductions of $H_c$ (Figure S5c) to find that the $H_c$ reduction by $I$(FGT)=2 mA is comparable to that by the temperature increase of 60 K. Note that it is considerably higher than the measured temperature increase due to the Joule heating, which is only 40 K. This exercise again indicates that there is a piece of clear experimental evidence for a more nontrivial effect in our data.

We successfully separated the $H_c$ reduction into the Joule heating contribution and the non-Joule-heating contribution in Figure S5d. At 2 mA, the Joule heating effect reduces $H_c$ from ~1517 to ~854 Oe, whereas the non-Joule-heating effect reduces $H_c$ from ~854 to ~534 Oe. Though the Joule heating effect is relatively more substantial for the sample S5, one should note that the reduction from ~854 to ~534 Oe by the non-Joule-heating effect is still a significant value when compared to the modulation in FePt from ~1160 to ~1110 Oe by the ionic liquid gating[2]. From such non-Joule-heating contribution, we estimated the strength of the spin-orbit field $|\mathbf{H}_{SOT}|$~$|\Gamma_0 J|$~51 Oe (Figure 3e) for $J$=1 mA/$\mu$m$^2$, which is two orders of magnitude larger than that produced by heavy metals such as Pt and Ta[3], and comparable to that produced by the topological insulator BiSb[4] (Figure 3f).

1.3 **COMSOL simulations for Joule heating (samples S1-S5).**

We also performed finite-element temperature simulations using the COMSOL package to analyze the Joule heating effect with the details given below. Figure S6a shows a schematic of the sample structure. Si and $SiO_2$ layers are 0.525 mm and 285 nm thick, respectively. Both layers are taken to be square-shaped with an area of 4 mm (length) × 4 mm (width). A cuboid-shaped FGT is located on top of the $SiO_2$ layer at the center of the $SiO_2$ layer. The geometry (length, width, and thickness) of the FGT layer is chosen to match with the actual values of FGT in the samples S1, S2, S3, S4, and S5. For the sample S4, which is not in the cuboid shape but instead more like a thin trigonal prism, it is also modeled as a cuboid for simplicity. Two Au electrodes are connected to FGT (Figure S6a). The width of the Au electrodes is chosen to match with that of FGT, and the length of the Au electrodes is set to 0.7 mm. The thickness of the Au electrodes is 95 nm. In the region where the Au electrodes are stacked over FGT, their interfaces are modeled as 3 nm-thick contact resistances (marked by red color in Figure S6a). The length of each stacked region is set to one-quarter of the FGT length so that the two contact resistances (two red-colored areas in Figure S6a) cover half of the FGT. The thickness of the Au electrodes in the stacked regions is chosen to be 95 nm – (FGT thickness) – (3 nm for the contact resistance).

The heat generation rate $j^2\rho$ is applied at each location, where $j$ and $\rho$ are the local current density and the local electrical resistivity, respectively. $j$ is nonzero only in the Au electrodes, the contact resistances, and FGT. For a given current $I$ injected to each sample, $j$ is set to $j=I/A$, where $A$ is the cross-sectional area. $A$ for each contact resistance is set to the area of each contact resistance, which is a quarter of the FGT area. This choice of $A$ for the contact resistances implies that the current flow from the Au electrodes to FGT occurs entirely through the contact resistances.

Since the helium gas surrounds the samples in actual experiments, the calculation also considers a possible heat exchange between the helium gas and the material layers, and even the heat convection in the helium gas. For this, the heat flux boundary condition is imposed on the outermost surfaces of the samples exposed to the helium gas. The helium gas temperature, located far away from the samples, is set to 2 K, and the heat transfer coefficient for the helium gas is set to 60 $W/m^2K$. The latter value is chosen so that the steady-state temperature at the side edge of the $SiO_2$ layer, which is 2 mm away from the FGT, agrees with the temperature measured in the sample S1 by using a CERNOX thermometer located 2 mm away from FGT. The same value of the heat transfer coefficient is used for the other samples as well.

To calculate the steady-state temperature, material parameters need to be specified, such as thermal conductivity, electrical resistivity, and specific heat. We adopted the material parameters taken after several previous reports for Si[5], $SiO_2$[6], gold[7], and FGT[8]. We used the value of the in-plane thermal conductivity $\kappa_{IP}$ taken from previous work for FGT[8]. Unfortunately, we were unable to find in the literature the value of the out-of-plane thermal conductivity $\kappa_{OP}$ for FGT. Therefore, we took a ratio of $\kappa_{OP} = 0.01\times\kappa_{IP}$ as a reasonable informed-guess, which is motivated by the observation that the ratio $\kappa_{OP}/\kappa_{IP}$ of about 0.01 was found for other 2D materials, including graphite[9]. To estimate how sensitive the FGT temperature is to $\kappa_{OP}$, we calculated the temperature of the FGT for the following two choices of the $\kappa_{OP}$ values; $\kappa_{OP} = \kappa_{IP}$ (Figure S6b) and $\kappa_{OP} = 0.01\times\kappa_{IP}$ (Figure S6c). For all the cases of samples and currents considered, the latter choice results in a more significant temperature increase. Still, the temperature difference from the former choice stays below 10 K. Although the temperature increase depends on the ratio $\kappa_{OP}/\kappa_{IP}$., this dependence is not that crucial. We thus conclude that the choice $\kappa_{OP} = 0.01\times\kappa_{IP}$ is sufficient for our purpose.

The contact resistance between the Au electrodes and FGT was also considered in the calculation by introducing a thin (3 nm) highly resistive region (marked by red color in Figure S6a). We chose the resistivity so that the resistance of each highly resistive region matches the experimental contact resistance. We verified that the calculated temperature is not sensitive to the thickness choice of the highly resistive areas.

Another important factor considered for the calculation is the temperature dependence of the material parameters. Most material parameters exhibit a modest temperature dependence (≤10 %). We tested whether their temperature dependences are essential for the steady-state temperature of the FGT. We found that their temperature dependence affects the steady-state temperature of the FGT only by a small amount (~2 K). We thus neglected the temperature dependence of those material parameters. To be more specific, we used most of the parameters taken at 10 K. The value of the electrical resistivity of FGT was set to its value at 2 K. For all the sample, the resistivity at 2 K is higher than the resistivity at higher temperatures (see Figure 1c and Figure S5b) up to the highest temperature achieved by the Joule heating. Thus the choice of the resistivity at 2 K tends to overestimate the Joule-heated temperature increase in our calculation.

An exception is the thermal conductivity of $SiO_2$, which is known to increase by about an order of magnitude when the temperature is raised from 2 to 30 K. Depending on which value of the thermal conductivity is used in the calculation, the steady-state temperature of FGT varies by more than 10 K. Thus the thermal conductivity of $SiO_2$ appears to be the most critical factor for the FGT temperature. We thus set the value of the thermal conductivity by imposing the self-consistency condition. We first set the value of the thermal conductivity to its value at a specific temperature $T^*$ and checked if the calculated steady-state temperature at the $SiO_2$ layer matches $T^*$. The same self-consistency condition was imposed for each current used in our calculation. The resulting steady-state temperature of FGT is shown in Figure S6b ($\kappa_{OP} = \kappa_{IP}$) and S6c ($\kappa_{OP} = 0.01\times\kappa_{IP}$) for various samples. The temperature dependence on the current is non-quadratic due to the temperature dependence of the $SiO_2$ thermal conductivity.

We finally compare the calculated temperatures (Figure S6c for $\kappa_{OP} = 0.01\times\kappa_{IP}$) with the experimentally assessed temperatures. For the sample S4: of which temperature is evaluated experimentally by using $NbSe_2$ as a nanofabricated thermometer (Figure S4), the calculated (experimentally assessed) temperatures are 11 (20) and 27 (24) K for the current of 1.0 and 2.0 mA, respectively. For the sample S5: of which temperature is experimentally assessed by using $R_{xx}$ as an internal thermometer (Figure S5), the calculated (experimentally assessed) temperatures are 21 (26) and 46 (40) K for the current of 1.0 and 2.0 mA, respectively. For both samples, the calculated temperatures are lower (by 5 to 9 K) than the experimentally assessed temperatures for the smaller current of 1.0 mA but higher (by 4 to 6 K) than the experimentally evaluated temperatures for the higher current of 2.0 mA. Note that the calculated and experimentally assessed temperatures agree with each other within 10 K, which validates the COMSOL simulations.

### 1.4 **Direct comparison of the temperature- and current-induced reductions of $H_c$ (samples S1 and S3).**

We can also use the measured temperature- and current-dependent Hall signals to examine whether the Joule heating can be the single primary explanation for the experimentally observed $H_c$ reduction for the samples S1. Figure S7a,b shows that $H_c$ for the sample S1 starts to decrease at 0.1 mA and reduces more significantly after 0.2 mA. In comparison, the $R_{xy}$-$H$ hysteresis loop measured at low current $I_{Sample}$=0.1 mA (see Figure S7c) shows little signs of $H_c$ change up to 10 K, and $H_c$ only starts to decrease from 20 K. Using these observations, we can make a direct comparison of the temperature- and the current-induced reductions of $H_c$ in Figure S7d. If the Joule heating were the sole source of the effects shown in Figure S7a, the FGT temperature should increase at least to ~50 and 100 K for the current of 0.2 and 1.0 mA, respectively. These temperatures are significantly higher than the estimate of 5 and 14 K for the corresponding situations obtained from our COMSOL calculations (see Figure S6c). According to the COMSOL calculations, the FGT temperature rises only to 23.5 K even for the highest current 1.5 mA when the sample temperature is set to 2 K. Therefore, we conclude that the Joule-heating contribution to the $H_c$ reduction cannot entirely explain the $I$-control of $H_c$ observed in the sample S1.

We can also make a similar analysis using the data of the sample S3, which is the thinnest sample and thus expected to have the highest current density by a factor of 3 as compared with all the other samples. For example, Figure S8a shows that $H_c$ is significantly reduced by increasing current at 2 K, from ~2220 Oe at 0.05 mA to ~30 Oe at 1.5 mA (Figure S8b). We also measured the $R_{xy}$-$H$ hysteresis loops with a small current 0.05 mA at various temperatures (Figure S2), from which the $H_c$-$T$ curve is also extracted (FigureS8c). Using these observations, we can make a direct comparison of the temperature- and the current-induced reductions of $H_c$ in Figure S8d. One important point to note is that the current of 1.5 mA shown in Figure S8a produces effects as strong as that by the temperature increase of ~125 K if the Joule heating ought to be the sole source of the $H_c$ reduction. Note that such temperature increase is far higher than the estimated 62 K from our COMSOL calculation (see Figure S6c). Considering that the Joule heating to 62 K reduces $H_c$ to ~400 Oe (Figure S8c,d), the non-Joule-heating effect is expected to further reduce $H_c$ from ~400 to ~30 Oe at 1.5 mA. Although this reduction of ~370 Oe is smaller than the total (including the Joule heating effect) reduction of ~2190 Oe (=2220-30), it is still much larger than the decrease of ~50 Oe (from 1160 to 1110 Oe) achieved for FePt[2] using the ionic gating. Thus we conclude that even for the thinnest sample S3, the non-Joule-heating effect of nontrivial origin induces a considerable reduction of $H_c$.

## 2. **Quantification of SOT by AMR.**

### 2.1 **Theoretical analysis of the current effect on AMR.**

#### 2.1-A **Longitudinal resistance.**

The longitudinal resistance $R_{x'x'}$ (current and voltage both along $x'$ direction) of FGT depends on the magnetization direction $\hat{\mathbf{m}}$ of FGT. This anisotropy of $R_{x'x'}$ or the anisotropic magnetoresistance (AMR) can be written as

$$R_{x'x'}\left(\hat{\mathbf{m}}\right) = R_{xx}\left(\hat{\mathbf{x}}'\right) m_{x'}^2 + R_{x'x'}\left(\hat{\mathbf{y}}'\right) m_{y'}^2 + R_{x'x'}\left(\hat{\mathbf{z}}'\right) m_{z'}^2 . \tag{S2-1}$$

Here it is assumed that FGT lies in the $x'y'$ plane. For $\hat{\mathbf{m}} = m_{x'}\hat{\mathbf{x}}' + m_{y'}\hat{\mathbf{y}}' + m_{z'}\hat{\mathbf{z}}' = \sin\theta'\cos\phi'\hat{\mathbf{x}}' + \sin\theta'\sin\phi'\hat{\mathbf{y}}' + \cos\theta'\hat{\mathbf{z}}'$, this equation can be written in a more convenient form,

$$R_{x'x'}\left(\hat{\mathbf{m}}\right) = R_{x'x'}\left(\hat{\mathbf{y}}'\right) + \left(\mathrm{OAMR}\right)\cos^2\theta' + \left(\mathrm{IAMR}\right)\sin^2\theta'\cos^2\phi' , \tag{S2-2}$$

where the in-plane AMR (IAMR) and the out-of-plane AMR (OAMR) are defined as

$$\left(\mathrm{IAMR}\right) \equiv R_{x'x'}\left(\hat{\mathbf{x}}'\right) - R_{x'x'}\left(\hat{\mathbf{y}}'\right) , \tag{S2-3}$$

$$\left(\mathrm{OAMR}\right) \equiv R_{x'x'}\left(\hat{\mathbf{z}}'\right) - R_{x'x'}\left(\hat{\mathbf{y}}'\right) . \tag{S2-4}$$

Here the direction of $\hat{\mathbf{m}}$ is determined from the free energy minimization.

2.1-B **Free energy minimization of FGT.**

The total free energy of FGT $f_{\mathrm{tot}}(\hat{\mathbf{m}})$ is given by

$$f_{\mathrm{tot}}(\hat{\mathbf{m}}) = -\frac{1}{2}M_{\mathrm{s}}K_z m_z^2 + M_{\mathrm{s}}\Gamma_0\left[J_y m_x m_y - \frac{1}{2}\left(m_x^2 - m_y^2\right)\right] - \mathbf{H}\cdot M_{\mathrm{s}}\hat{\mathbf{m}}, \tag{S2-5}$$

where $x$ is the armchair direction, and $y$ is the zigzag direction within FGT. For the $x'$, $y'$, $z'$ directions introduced above: the $x'$ direction is defined as the direction of the current of $\mathbf{J} = J\hat{\mathbf{x}}'$, the $x$, $y$, $z$ directions are related to the $x'$, $y'$, $z'$ directions in the following way. The $z$-direction is the same as the $z'$ direction, but the $x$ and $y$ axes can be rotated from the $x'$ and $y'$ axes by the angle $\phi_J$. That is,

$$\mathbf{J} = J_x\hat{\mathbf{x}} + J_y\hat{\mathbf{y}} = J\hat{\mathbf{x}}' \;\Rightarrow\; J_x = J\cos\phi_J,\; J_y = J\sin\phi_J$$

$$m_{x'} = m_x\cos\phi_J + m_y\sin\phi_J,\; m_{y'} = -m_x\sin\phi_J + m_y\cos\phi_J,\; m_{z'} = m_z. \tag{S2-6}$$

Using the parameterization, $\hat{\mathbf{m}} = \sin\theta\cos\phi\hat{\mathbf{x}} + \sin\theta\sin\phi\hat{\mathbf{y}} + \cos\theta\hat{\mathbf{z}}$, $f_{\mathrm{tot}}(\hat{\mathbf{m}})$ reads

$$\begin{aligned} f_{\mathrm{tot}}(\hat{\mathbf{m}}) = -\frac{M_{\mathrm{s}}}{2}\Big\{&K_z\cos^2\theta + \Gamma_0 J\sin^2\theta\cos\left(2\phi+\phi_J\right) \\ &+ 2H\left[\sin\theta\sin\theta_{\mathrm{H}}\cos\left(\phi-\phi_{\mathrm{H}}\right) + \cos\theta\cos\theta_{\mathrm{H}}\right]\Big\} \end{aligned}, \tag{S2-7}$$

where $\mathbf{H} = H\left(\sin\theta_{\mathrm{H}}\cos\phi_{\mathrm{H}}\hat{\mathbf{x}} + \sin\theta_{\mathrm{H}}\sin\phi_{\mathrm{H}}\hat{\mathbf{y}} + \cos\theta_{\mathrm{H}}\hat{\mathbf{z}}\right)$ is used.

The minimization of $f_{\mathrm{tot}}(\hat{\mathbf{m}})$ for $\hat{\mathbf{m}}$ is, in principle, straightforward but quite tedious. For the minimization, we assume that $\theta_H$ is close to $\pi/2$, and we regard $\delta\theta_H \equiv \theta_H - \pi/2$ and $J$ as small expansion parameters. This way, one finds that $f_{\mathrm{tot}}(\hat{\mathbf{m}})$ is minimized when $\phi$ and $\theta$ have the following values,

$$\phi = \phi_H - \frac{\Gamma_0 J}{H}\sin\left(2\phi_H + \phi_J\right) + \left(\frac{\Gamma_0 J}{H}\right)^2\sin\left(4\phi_H + 2\phi_J\right) + O\left(\delta\theta_H, J\right)^3, \tag{S2-8}$$

$$
\begin{aligned}
\theta = & \frac{\pi}{2} + \delta\theta_H \frac{H}{H-K_z} - \frac{1}{6}\left(\delta\theta_H\right)^3 \frac{K_z H}{\left(H-K_z\right)^2} \frac{4H^2 - K_z^2}{\left(H-K_z\right)^2} \\
& - \delta\theta_H \frac{\Gamma_0 J}{H-K_z} \frac{H}{H-K_z} \cos\left(2\phi_H + \phi_J\right) \\
& - \delta\theta_H \left(\frac{\Gamma_0 J}{H-K_z}\right)^2 \left[\frac{3}{2}\sin^2\left(2\phi_H + \phi_J\right) - \frac{H}{H-K_z}\cos^2\left(2\phi_H + \phi_J\right)\right] \\
& + O\left(\delta\theta_H, J\right)^4
\end{aligned} \quad . \qquad \text{(S2-9)}
$$

2.1-C **Geometry.**

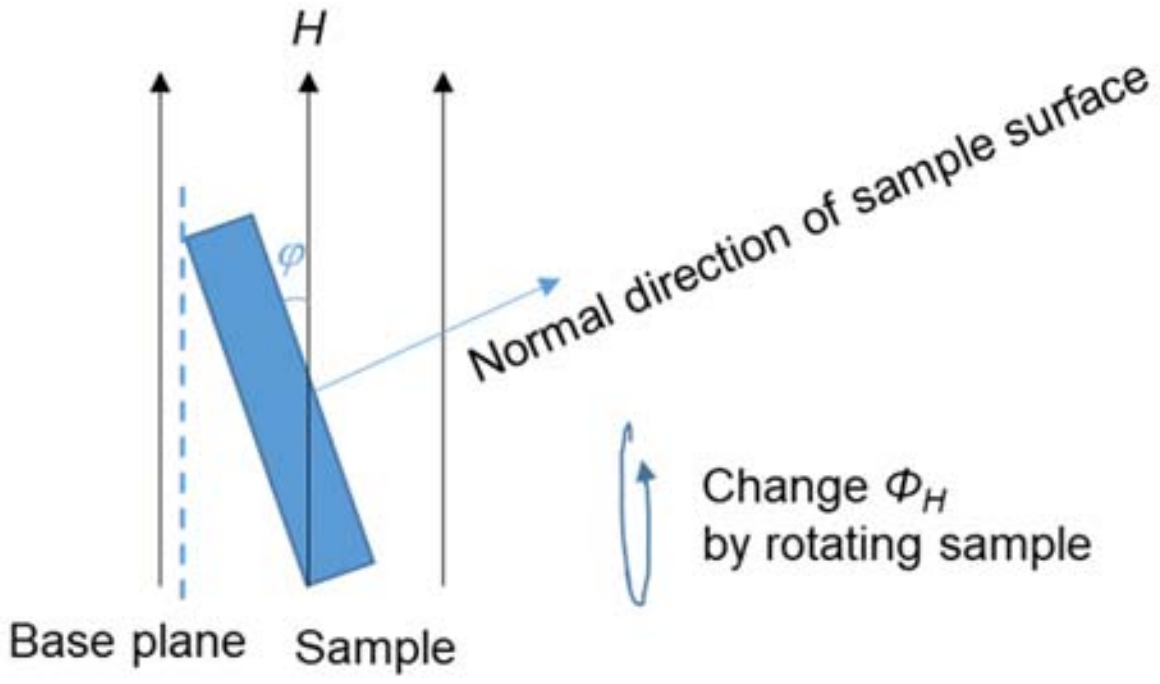

We must consider other experimental factors such as the misalignment of the samples to achieve a more realistic calculation further. We took the geometry as sketched above, where the FGT sample plane is misaligned from the base plane by the angle $\varphi$. Although the misalignment is not intended, it always exists in experiments and turns out to be essential for the analysis of experimental data. To be more specific, we take the "Normal direction of FGT surface" as the $z$-direction and assume that the direction of $\mathbf{J}$ deviates by the angle $\phi_0$ from the cross-section of the FGT plane and the paper plane. We also assume that the magnetic field $\mathbf{H}$ is rotated around the normal direction of the base plane (in reality, $\mathbf{H}$ is fixed, and the base plane together with FGT is rotated around the base plane normal direction). Then $\delta\theta_H = \theta_H - \pi/2$ varies as follows during the rotation of $\mathbf{H}$,

$$
\sin\delta\theta_H = \sin\varphi\cos\left(\phi_H - \phi_0\right). \qquad \text{(S2-10)}
$$

$\delta\theta_H$ does depend on $\phi_H$. For this reason, not only (IAMR) but also the (OAMR) affects $\phi_H$-

dependence of $R_{x'x'}(\hat{\mathbf{m}})$ as shown later. Recalling that $\delta\theta_H$ is a small parameter, $\varphi$ is also a small parameter.

2.1-D $\phi_H$**-dependence of** $R_{x'x'}(\hat{\mathbf{m}})$**.**

When $\hat{\mathbf{m}}$ is determined from the free energy minimization, $R_{x'x'}(\hat{\mathbf{m}})$ depends on $\phi_H$ in two ways; through (IAMR) and (OAMR). For (IAMR), we find

$$\begin{aligned}\sin^2\theta' &= \sin^2\theta \\ &= 1-\left(\frac{H}{H-K_z}\right)^2 \sin^2\varphi\cos^2\left(\phi_H-\phi_0\right)+O\left(\varphi,J\right)^3,\end{aligned} \qquad \text{(S2-11)}$$

$$\begin{aligned}&\cos^2\phi' = \cos^2\left(\phi-\phi_J\right) \\ &= \cos^2\left(\phi_H-\phi_J\right)+\frac{\Gamma_0 J}{H}\sin\left(2\phi_H-2\phi_J\right)\sin\left(2\phi_H+\phi_J\right) \\ &\quad +\left(\frac{\Gamma_0 J}{H}\right)^2\left[\sin\left(2\phi_H-2\phi_J\right)\sin\left(4\phi_H+2\phi_J\right)-\cos\left(2\phi_H-2\phi_J\right)\sin^2\left(2\phi_H+\phi_J\right)\right] \\ &\quad +O\left(\varphi,J\right)^3\end{aligned}. \qquad \text{(S2-12)}$$

For (OAMR), we find

$$\begin{aligned}&\cos^2\theta' = \cos^2\theta \\ &= \sin^2\varphi\cos^2\left(\phi_H-\phi_0\right)\cdot\left(\frac{H}{H-K_z}\right)^2 \\ &\quad\times\Bigg\{1-\sin^2\varphi\cos^2\left(\phi_H-\phi_0\right)\cdot\frac{HK_z\left(2H-K_z\right)}{\left(H-K_z\right)^3}-\frac{2\Gamma_0 J}{H-K_z}\cos\left(2\phi_H+\phi_J\right) \\ &\qquad -\frac{3\left(\Gamma_0 J\right)^2}{H\left(H-K_z\right)}\left[\sin^2\left(2\phi_H+\phi_J\right)-\frac{H}{H-K_z}\cos^2\left(2\phi_H+\phi_J\right)\right]+O\left(\varphi,J\right)^3\Bigg\}\end{aligned}. \qquad \text{(S2-13)}$$

We are now ready to examine the $\phi_H$-dependence of $R_{x'x'}(\hat{\mathbf{m}})$,

$$R_{x'x'}(\hat{\mathbf{m}})-R_{x'x'}(\hat{\mathbf{x}})=\left(\text{IAMR}\right)\sin^2\theta\cos^2\left(\phi-\phi_J\right)+\left(\text{OAMR}\right)\cos^2\theta. \qquad \text{(S2-14)}$$

Since both $\phi$ and $\theta$ depend on $\phi_H$, both (IAMR) and (OAMR) generate the $\phi_H$-dependence of $R_{x'x'}(\hat{\mathbf{m}})$. Here we present its expression for the case $\phi_J = 0$ and $\phi_0 = \pi/2$ since this choice results in a good fitting of our experimental result. From the (IAMR), one obtains

$$\begin{aligned}
&(\mathrm{IAMR})\sin^2\theta\cos^2(\phi-\phi_J) \\
&=(\mathrm{IAMR})\cos^2\phi_H\left\{1-\left(\frac{H}{H-K_z}\right)^2\sin^2\varphi\sin^2\phi_H\right. \\
&\qquad\left.+\frac{4\Gamma_0 J}{H}\sin^2\phi_H+4\left(\frac{\Gamma_0 J}{H}\right)^2\sin^2\phi_H\cos 2\phi_H+O(\varphi,J)^3\right\}
\end{aligned}, \tag{S2-15}$$

and from the (OAMR), one obtains

$$\begin{aligned}
&(\mathrm{OAMR})\cos^2\theta \\
&=(\text{R-OAMR})\sin^2\phi_H\left\{1-\frac{HK_z(2H-K_z)}{(H-K_z)^3}\sin^2\varphi\sin^2\phi_H\right. \\
&\left.-\frac{2\Gamma_0 J}{H-K_z}\cos 2\phi_H-\frac{3(\Gamma_0 J)^2}{H(H-K_z)}\left(\sin^2 2\phi_H-\frac{H}{H-K_z}\cos^2 2\phi_H\right)+O(\varphi,J)^3\right\}
\end{aligned}, \tag{S2-16}$$

where the renormalized OAMR (R-OAMR) is defined by

$$(\text{R-OAMR})\equiv(\mathrm{OAMR})\left(\frac{H}{H-K_z}\right)^2\sin^2\varphi . \tag{S2-17}$$

Before proceeding further, it is worth comparing the values of (IAMR) and (R-OAMR). For (IAMR)/(OAMR)=1/20 , $H$=9 T, $K_z$=4 T, $\varphi=\pi(5°/180°)$ (corresponding to 5°), one obtains

$$\frac{(\text{R-OAMR})}{(\mathrm{IAMR})}\equiv 20\times\left(\frac{9}{9-4}\right)^2\sin^2\frac{5\pi}{180}=0.49 . \tag{S2-18}$$

For (IAMR)/(OAMR)=1/20, $H$=9 T, $K_z$=4 T, $\varphi=\pi(7°/180°)$ (corresponding to 7°), one obtains

$$\frac{(\text{R-OAMR})}{(\mathrm{IAMR})}\equiv 20\times\left(\frac{9}{9-4}\right)^2\sin^2\frac{7\pi}{180}=0.96 . \tag{S2-19}$$

Thus for our experiments, (R-OAMR) is similar to (IAMR) in magnitude, emphasizing the importance of (R-OAMR).

When the (IAMR) contribution [Eq. (S2-15)] and the (R-OAMR) contribution [Eq. (S2-16)] are combined, one obtains after some algebra,

$$
\begin{aligned}
&R_{x'x'}(\hat{\mathbf{m}}) - R_{x'x'}(\hat{\mathbf{y}}') \\
&= C_0 + \cos^2\phi_H \left[ C_1 + \sin^2\phi_H \left( C_2 \cos^2\phi_H + C_3 \right) \right],
\end{aligned} \tag{S2-20}
$$

where the $\phi_H$ -independent contribution $C_0$ is given by

$$
C_0 = (\text{R-OAMR}) \left[ 1 - \frac{HK_z(2H-K_z)}{(H-K_z)^3} \sin^2\varphi + \frac{2\Gamma_0 J}{H-K_z} + \frac{3(\Gamma_0 J)^2}{(H-K_z)^2} + O(\varphi, J)^3 \right], \tag{S2-21}
$$

and the amplitudes $C_1$, $C_2$, $C_3$ of the $\phi_H$ -dependent contributions are given by

$$
\begin{aligned}
C_1 &= (\text{IAMR})\left[ 1 + O(\varphi, J)^3 \right] \\
&\quad - (\text{R-OAMR}) \left[ 1 - \frac{HK_z(2H-K_z)}{(H-K_z)^3} \sin^2\varphi + \frac{2\Gamma_0 J}{H-K_z} + \frac{3(\Gamma_0 J)^2}{(H-K_z)^2} + O(\varphi, J)^3 \right],
\end{aligned} \tag{S2-22}
$$

$$
C_2 = (\text{IAMR})\, 8\left( \frac{\Gamma_0 J}{H} \right)^2 + (\text{R-OAMR}) \frac{12(12H-K_z)(\Gamma_0 J)^2}{H(H-K_z)^2}, \tag{S2-23}
$$

$$
\begin{aligned}
C_3 &= (\text{IAMR}) \left[ -\sin^2\varphi \left( \frac{H}{H-K_z} \right)^2 + \frac{4\Gamma_0 J}{H} - 4\left( \frac{\Gamma_0 J}{H} \right)^2 \right] \\
&\quad + (\text{R-OAMR}) \left[ \sin^2\varphi \frac{HK_z(2H-K_z)}{(H-K_z)^3} - \frac{4\Gamma_0 J}{H-K_z} - \frac{12(12H-K_z)(\Gamma_0 J)^2}{H(H-K_z)^2} \right].
\end{aligned} \tag{S2-24}
$$

2.1-E **From a single-layer FGT to a multilayer FGT.**

Let the resistances of single A and B layers be $R_{x'x'}^{\mathrm{A}}$ and $R_{x'x'}^{\mathrm{B}}$, respectively. Then the resistance $R_{x'x'}^{\mathrm{multi}}$ of a multilayer FGT, where A and B layers are repeated $N$ times, is given by

$$
R_{x'x'}^{\mathrm{multi}} = \frac{1}{N} \left( \frac{1}{R_{x'x'}^{\mathrm{A}}} + \frac{1}{R_{x'x'}^{\mathrm{B}}} \right)^{-1} = \frac{1}{N} \frac{R_{x'x'}^{\mathrm{A}} R_{x'x'}^{\mathrm{B}}}{R_{x'x'}^{\mathrm{A}} + R_{x'x'}^{\mathrm{B}}}. \qquad \text{(S2-25).}
$$

When $R_{x'x'}^{\mathrm{A}}$ and $R_{x'x'}^{\mathrm{B}}$ differ from each other only by a tiny amount, which is indeed the case, Eq. (S2-25) can be approximated as

$$
R_{x'x'}^{\mathrm{multi}} = \frac{1}{N} \frac{R_{x'x'}^{\mathrm{A}} R_{x'x'}^{\mathrm{B}}}{R_{x'x'}^{\mathrm{A}} + R_{x'x'}^{\mathrm{B}}} \approx \frac{1}{2N} \frac{R_{x'x'}^{\mathrm{A}} + R_{x'x'}^{\mathrm{B}}}{2}. \tag{S2-26}
$$

Now let us evaluate the $\phi_H$-dependence of $R_{x'x'}^{\text{multi}}$. Together with Eq. (S2-26), one then obtains

$$2NR_{x'x'}^{\text{multi}}\left(\hat{\mathbf{m}}^{\text{A}},\hat{\mathbf{m}}^{\text{B}}\right)-\frac{R_{x'x'}^{\text{A}}\left(\hat{\mathbf{y}}'\right)+R_{x'x'}^{\text{B}}\left(\hat{\mathbf{y}}'\right)}{2} = C_0^{\text{multi}}+\cos^2\phi_H\left[C_1^{\text{multi}}+\sin^2\phi_H\left(C_2^{\text{multi}}\cos^2\phi_H+C_3^{\text{multi}}\sin^2\phi_H\right)\right]. \quad \text{(S2-27)}$$

The coefficients in Eq. (S2-27) are given by

$$C_0^{\text{multi}}=\frac{C_0^{\text{A}}+C_0^{\text{B}}}{2}=\left(\text{R-OAMR}\right)\left[1-\frac{HK_z\left(2H-K_z\right)}{\left(H-K_z\right)^3}\sin^2\varphi+\frac{3\left(\Gamma_0J\right)^2}{\left(H-K_z\right)^2}+O\left(\varphi,J\right)^3\right], \quad \text{(S2-28)}$$

$$C_1^{\text{multi}}=\frac{C_1^{\text{A}}+C_1^{\text{B}}}{2}=\left(\text{IAMR}\right)\left[1+O\left(\varphi,J\right)^3\right] -\left(\text{R-OAMR}\right)\left[1-\frac{HK_z\left(2H-K_z\right)}{\left(H-K_z\right)^3}\sin^2\varphi+\frac{3\left(\Gamma_0J\right)^2}{\left(H-K_z\right)^2}+O\left(\varphi,J\right)^3\right], \quad \text{(S2-29)}$$

$$C_2^{\text{multi}}=\frac{C_2^{\text{A}}+C_2^{\text{B}}}{2}=\left(\text{IAMR}\right)8\left(\frac{\Gamma_0J}{H}\right)^2+\left(\text{R-OAMR}\right)\frac{12\left(12H-K_z\right)\left(\Gamma_0J\right)^2}{H\left(H-K_z\right)^2}, \quad \text{(S2-30)}$$

$$C_3^{\text{multi}}=\frac{C_3^{\text{A}}+C_3^{\text{B}}}{2}=\left(\text{IAMR}\right)\left[-\sin^2\varphi\left(\frac{H}{H-K_z}\right)^2-4\left(\frac{\Gamma_0J}{H}\right)^2\right] +\left(\text{R-OAMR}\right)\left[\sin^2\varphi\frac{HK_z\left(2H-K_z\right)}{\left(H-K_z\right)^3}-\frac{12\left(12H-K_z\right)\left(\Gamma_0J\right)^2}{H\left(H-K_z\right)^2}\right]. \quad \text{(S2-31)}$$

Note that the following expression governs the $\phi_H$-dependence of $R_{x'x'}^{\text{multi}}\left(\hat{\mathbf{m}}^{\text{A}},\hat{\mathbf{m}}^{\text{B}}\right)$:

$$\cos^2\phi_H\cdot\left[1+\sin^2\phi_H\left(\frac{C_2^{\text{multi}}}{C_1^{\text{multi}}}\cos^2\phi_H+\frac{C_3^{\text{multi}}}{C_1^{\text{multi}}}\sin^2\phi_H\right)\right]. \quad \text{(S2-32)}$$

In accessing the $\phi_H$-dependence of $R_{x'x'}^{\text{multi}}\left(\hat{\mathbf{m}}^{\text{A}},\hat{\mathbf{m}}^{\text{B}}\right)$ against the so-called "standard" dependence of $\cos^2\phi_H$, we used the two ratios, $C_2^{\text{multi}}/C_1^{\text{multi}}$ and $C_3^{\text{multi}}/C_1^{\text{multi}}$.

2.2 **AMR measurement.**

The angular dependence of the normalized longitudinal resistance was measured with an in-plane magnetic field of $H$=9 T. Figure S10 shows the longitudinal resistance as a function of $\phi_H$ for the current of 0.75 mA **(a,d)**, 1.0 mA **(b,e)**, 1.25 mA **(c,f)**. The solid black lines in **(a,b,c)** denote the curve of the "conventional" angular dependence $\cos^2\phi_H$, which roughly works for the relatively small current of 0.75 mA. But the fitting becomes less satisfactory for the larger currents of 1.0 and 1.25 mA since the peaks become sharper in the experimental data than that predicted by the "conventional" angular dependence $\cos^2\phi_H$. On the other hand, the solid black lines in **(d,e,f)** denote the fitting results by the multilayer FGT AMR formula (S2-32) in Sec. 2.1. This formula makes the better fitting of the experimental data, including the sharper peaks. Note that the formula contains three fitting parameters ($\Gamma_0$, $\varphi$, OAMR/IAMR), and all three sets of data for the three currents are fitted by one set of parameters. Other independent measurements were used to determine all the other parameters in the fitting formula. Through this fitting, we obtained $|\Gamma_0|$=119 ± 28 Oe/(mA/μm$^2$), $\varphi$=5.59° ± 1.94°, and OAMR/IAMR=34.2 ± 17.4.

Here we clarify that the fitted values of the three parameters are reasonable based on our experiments (Figure S11) and the previous reports. Figure S11a,b shows the typical magnetoresistance ratio ($R$($H$)-$R$(0))/$R$(0) of FGT as a function of magnetic field $H$ with $H$ applied along $x$, $y$, $z$ directions, from which the OAMR/IAMR of FGT is estimated to be ~48 at 2 K and 9 T (Figure S11c). Such OAMR/IAMR roughly agrees with our fitted ratio of ~34.2 ± 17.4 by the AMR data, and with the estimated ratio of ~20 at 100 K and 5 T in the previous report[10]. In practical experiments with an applied in-plane magnetic field, it is difficult to achieve a perfect alignment as one cannot avoid an unwanted slight misalignment of samples to the field direction[10,11]. By comparing Figure S11d with the angle-dependent $R_{xy}$-$H$ curves in the Extended Data Figure 6 of a previous report[12], the misalignment angle $\varphi$ in our AMR experiment should be close to 5°, which is compatible with our fitted $\varphi$=5.59° ± 1.94° by the AMR data. Moreover, the value of $|\Gamma_0|$=119 Oe/(mA/μm$^2$) is on the same order of magnitude as that estimated from the $H_c$ reduction effect ($|\Gamma_0|\approx$50 Oe/(mA/μm$^2$)). Thus the values of all fitting parameters ($\varphi$, OAMR/IAMR, and $|\Gamma_0|$) are reasonable.

Finally, we comment on the possible effects of the Joule heating on the AMR analysis. Joule heating can affect the material parameters. For instance, the ratio OAMR/IAMR is 48 at 2 K (Figure S11c) but reduces as the temperature increases (c.f. 20 at 100 K[10]). Thus the Joule heating can introduce the current dependence of the ratio OAMR/IAMR. Considering that the AMR analysis in the preceding section takes into account the current dependence only through $|\Gamma_0 J|$, it is in principle possible that the current dependence of the ratio may be mixed with the current dependence of $|\Gamma_0 J|$, leading to the overestimation of $|\Gamma_0 J|$. However, if we take 48 and 20 as the upper and lower bounds of the ratio, both limits lie within the error range of the fitted ratio of 34.2 ± 17.4. It implies that the correction to $|\Gamma_0|$ due to the current dependence of the ratio can be estimated by the error bar (±28) of the fitted value of $|\Gamma_0|$=119 ± 28 Oe/(mA/μm$^2$). Thus we conclude that although the fitted value of $|\Gamma_0|$ may need modification (a few tens of percent) due to the Joule heating effect, the current-dependent variation and deviation from "conventional" $\cos^2\phi_H$ behavior of AMR cannot be explained by the Joule heating, providing a piece of robust evidence in support of the current-induced SOT.

3. **Theoretical calculations of $\Gamma_0$.**

As another essential independent check of the nontrivial nature of the observed reduction of $H_c$, we calculated $\Gamma_0$ for a multilayer FGT. The field-dependent electronic structures are obtained using the density functional theory (DFT) within the pseudopotential plane-wave basis implemented in the Quantum ESPRESSO package[13]. For exchange-correlation functionals, we used the Perdew-Burke-Ernzerhof parameterization[14] with the Kresse-Joubert projected augmented wave method (KJPAW)[15]. Spin-orbit interaction is included via the second-order perturbation approach. The kinetic energy cutoffs for the wave function and charge density and potential are set to 60 and 250 Ry, respectively; **k** point mesh of 16×16×4 is used. The directions of the magnetic moments of Fe atoms are constrained in [100], [010], [001] directions for the field dependence of the band structures.

Based on this electronic structure, we use the WANNIER90 program to obtain the maximally localized Wannier functions (MLWFs)[16] consisting of $d_{xy}, d_{yz}, d_{zx}, d_{x^2-y^2}, d_{z^2}$ orbitals for four Fe III atoms and two Fe II atoms, and $p_x, p_y, p_z$ orbitals for two Ge atoms and four Te atoms (total of 96 MLWFs (48 orbitals times two spin directions)). With these MLWFs, we construct a $96\times96$ tight-binding Hamiltonian $H(\mathbf{k})$. The spin ($\mathbf{S}$) and orbital ($\mathbf{L}$) angular momentum operators are locally defined in an atomic orbital representation.

We then evaluated the current-induced magnetic moment $\delta\boldsymbol{\mu}$ at Fe atoms since the magnetism in FGT arises from Fe atoms. For an external electric field $\mathbf{E}=E_j\hat{\mathbf{j}}$ and the magnetization $\mathbf{m}=m_k\hat{\mathbf{k}}$ ($j,k=x,y,z$), $\delta\boldsymbol{\mu}$ at the Fe atom $\alpha$ is evaluated by employing the Kubo formula,

$$\begin{aligned}&\delta\boldsymbol{\mu}(\mathrm{Fe}_\alpha)\\&=e\hbar E_j V\sum_{n,m}\int\frac{d^3\mathbf{k}}{(2\pi)^3}(f_{n\mathbf{k}}-f_{m\mathbf{k}})\,\mathrm{Im}\left[\frac{\langle u_{n\mathbf{k}}|P_\alpha\boldsymbol{\mu}P_\alpha|u_{m\mathbf{k}}\rangle\langle u_{m\mathbf{k}}|v_j(\mathbf{k})|u_{n\mathbf{k}}\rangle}{(E_{n\mathbf{k}}-E_{m\mathbf{k}}+i\eta)^2}\right],\end{aligned}\tag{S3.1}$$

where the magnetic moment operator $\boldsymbol{\mu}$ is defined by $\boldsymbol{\mu}=\gamma(2\mathbf{S}+\mathbf{L})$, $\gamma(<0)$ is the electron gyromagnetic ratio, $P_\alpha$ is the projection operator to the Fe atom $\alpha$, $e>0$ is the elementary charge, $V$ is the unit cell volume, $f_{n\mathbf{k}}$ is the Fermi-Dirac distribution function, $v_j(\mathbf{k})=(1/\hbar)(\partial H(\mathbf{k})/\partial k_j)$ is the velocity operator along the $j$ direction, $|u_{n\mathbf{k}}\rangle$ is the cell-periodic part of the Bloch state, and $E_{n\mathbf{k}}$ is the corresponding energy eigenvalue. Here the spectral broadening $\eta$ is adopted to be 2.5 meV, which corresponds to the temperature of ~30 K. The **k**-integration is carried out on the uniformly distributed $100\times100\times100$ **k**-point mesh, where the integration is converged sufficiently.

$\delta\boldsymbol{\mu}(\mathrm{Fe}_\alpha)$ is related to the torque $\boldsymbol{\tau}_\alpha$ at the Fe atom $\alpha$ as follows,

$$\boldsymbol{\tau}_\alpha=-\frac{J_{\mathrm{ex}}}{|\gamma|\hbar^2}\mathbf{m}\times\delta\boldsymbol{\mu}(\mathrm{Fe}_\alpha),\tag{S3.2}$$

where $J_{\text{ex}}$ is the strength of the exchange interaction. Compared with the conventional expression for the torque $\boldsymbol{\tau}_\alpha = -|\gamma| \mathbf{m} \times \mathbf{H}_{\text{SOT}}$, one finds

$$|\gamma| \mathbf{H}_{\text{SOT}} = \frac{J_{\text{ex}}}{|\gamma| \hbar^2} \delta \boldsymbol{\mu} (\text{Fe}_\alpha). \tag{S3.3}$$

In a *single*-layer FGT, which contains three Fe atoms within its unit cell, its symmetries constraint $\mathbf{H}_{\text{SOT}}$ [17] to have the following form,

$$\mathbf{H}_{\text{SOT}} = \Gamma_0 \left[ \left( m_x J_x - m_y J_y \right) \hat{\mathbf{x}} - \left( m_y J_x + m_x J_y \right) \hat{\mathbf{y}} \right]. \tag{S3.4}$$

Here we remark that a single-layer FGT contains three Fe atoms (1 Fe II atoms and 2 Fe III atoms) in a unit cell. The symmetries apply to the set of the three Fe atoms rather than each Fe atom (each Fe III atom does not satisfy the symmetries). Thus $\mathbf{H}_{\text{SOT}}$ in Eq. (S3.4) is the effective field *averaged* over the three Fe atoms. In a *multi*layer FGT, where individual FGT layers make AB-type stacking[18] (Figure S12), the same symmetries apply to each A-type or B-type layer FGT. Thus the effective magnetic fields $\mathbf{H}^{\text{A}}_{\text{SOT}}$ (averaged over three Fe atoms in an A-type layer) and $\mathbf{H}^{\text{B}}_{\text{SOT}}$ (averaged over three Fe atoms in a B-type layer) also have the structure in Eq. (S3.4). However, there is one difference. Since a B-type layer is rotated 180° from an A-type layer with respect to the *z*-axis, $\Gamma_0$'s for an A-type ($\Gamma^{\text{A}}_0$) and B-type ($\Gamma^{\text{B}}_0$) layer have the opposite signs, $\Gamma^{\text{A}}_0 = -\Gamma^{\text{B}}_0 = \Gamma_0$;

$$\mathbf{H}^{\text{A}}_{\text{SOT}} = \Gamma_0 \left[ \left( m_x J_x - m_y J_y \right) \hat{\mathbf{x}} - \left( m_y J_x + m_x J_y \right) \hat{\mathbf{y}} \right], \tag{S3.5}$$

$$\mathbf{H}^{\text{B}}_{\text{SOT}} = -\Gamma_0 \left[ \left( m_x J_x - m_y J_y \right) \hat{\mathbf{x}} - \left( m_y J_x + m_x J_y \right) \hat{\mathbf{y}} \right]. \tag{S3.6}$$

Figure S13 shows $\delta \boldsymbol{\mu}^{\text{A}} / \gamma E_y$ for $\mathbf{E} = E_y \hat{\mathbf{y}}$, where $\delta \boldsymbol{\mu}^{\text{A}}$ denotes the total magnetic moment summed over three Fe atoms in an A-type layer. For $\mathbf{m} = \hat{\mathbf{x}}$ ($\mathbf{m} = \hat{\mathbf{y}}$), we find $\delta \boldsymbol{\mu}^{\text{A}}$ to be almost parallel to $\hat{\mathbf{y}}$ ($\hat{\mathbf{x}}$). Figure S13a,b shows $\delta\boldsymbol{\mu}^{\mathrm{A}}\cdot\hat{\mathbf{y}}/\gamma E_y$ and $\delta\boldsymbol{\mu}^{\mathrm{A}}\cdot\hat{\mathbf{x}}/\gamma E_y$ for $\mathbf{m}=\hat{\mathbf{x}}$ and $\mathbf{m}=\hat{\mathbf{y}}$, respectively. According to Eqs. (S3.3) and (S3.5), the relation $\delta\boldsymbol{\mu}^{\mathrm{A}}\left(\mathbf{m}=\hat{\mathbf{y}}\right)\cdot\hat{\mathbf{x}}=\delta\boldsymbol{\mu}^{\mathrm{A}}\left(\mathbf{m}=\hat{\mathbf{x}}\right)\cdot\hat{\mathbf{y}}$ should be satisfied. The calculation results shown in Figure S13a,b satisfy the relation well for all values of the Fermi energy $E_{\mathrm{F}}$, consistent with the prediction (S3.5) based on the symmetries. We also verify that the total magnetic moment $\delta\boldsymbol{\mu}^{\mathrm{B}}$ summed over three Fe atoms in a B-type layer (not shown) satisfies $\delta\boldsymbol{\mu}^{\mathrm{B}}=-\delta\boldsymbol{\mu}^{\mathrm{A}}$ consistent with Eqs. (S3.5) and (S3.6).

Now we can evaluate the value of $\Gamma_0$ using the following formula;

$$\Gamma_0=-\frac{\rho}{3}\frac{J_{\mathrm{ex}}}{|\gamma|\hbar}\frac{\delta\boldsymbol{\mu}^{\mathrm{A}}\left(\mathbf{m}=\hat{\mathbf{x}}\right)\cdot\hat{\mathbf{y}}}{\gamma\hbar E_y}, \tag{S3.7}$$

where $\rho$ is the longitudinal resistivity of a multilayer FGT, which is $\approx 10^{-5}$ Ω·m and the exchange interaction $J_{\mathrm{ex}}\approx 1.4$ eV is estimated from the spin splitting of the DFT band structure. Compared with Eq. (S3.3), there is an extra factor 1/3 in Eq. (S3.7). This factor is introduced to take account of the fact that $\delta\boldsymbol{\mu}^{\mathrm{A}}$ is the total magnetic moment of the three Fe atoms in a type-A layer instead of the averaged magnetic moment. From Eq. (S3.7) and Figure S13a, we obtain the following value of $\Gamma_0$;

(1) At $E_{\mathrm{F}}=0$ eV,

$$\frac{\delta\boldsymbol{\mu}^{\mathrm{A}}\left(\mathbf{m}=\hat{\mathbf{x}}\right)\cdot\mathbf{y}}{\gamma\hbar E_y}\approx 4\times 10^{-12}\ \left(\mathrm{V/m}\right)^{-1}, \tag{S3.8}$$

which corresponds to

$$\Gamma_0\approx 3\times 10^{-13}\ \mathrm{T/(A/m^2)}=300\ \mathrm{Oe}/(10^{11}\ \mathrm{A/m^2})=3\ \mathrm{Oe/(mA/\mu m^2)}. \tag{S3.9}$$

(2) At $E_{\mathrm{F}}=+0.14$ eV,

$$\frac{\delta \boldsymbol{\mu}^{\mathrm{A}}\left(\mathbf{m}=\hat{\mathbf{x}}\right)\cdot\mathbf{y}}{\gamma\hbar E_y} \approx -3\times10^{-11}\ \left(\mathrm{V/m}\right)^{-1}, \tag{S3.10}$$

which corresponds to

$$\Gamma_0 \approx -2.3\times10^{-12}\ \mathrm{T/(A/m^2)} = -2300\ \mathrm{Oe/(10^{11}\ A/m^2)} = -23\ \mathrm{Oe/(mA/\mu m^2)}. \tag{S3.11}$$

(3) At $E_{\mathrm{F}}$ = –0.36 eV,

$$\frac{\delta \boldsymbol{\mu}^{\mathrm{A}}\left(\mathbf{m}=\hat{\mathbf{x}}\right)\cdot\mathbf{y}}{\gamma\hbar E_y} \approx 4\times10^{-11}\ \left(\mathrm{V/m}\right)^{-1}, \tag{S3.12}$$

which corresponds to

$$\Gamma_0 \approx 3\times10^{-12}\ \mathrm{T/(A/m^2)} = 3000\ \mathrm{Oe/(10^{11}\ A/m^2)} = 30\ \mathrm{Oe/(mA/\mu m^2)}. \tag{S3.13}$$

(4) At $E_{\mathrm{F}}$ = –0.56 eV,

$$\frac{\delta \boldsymbol{\mu}^{\mathrm{A}}\left(\mathbf{m}=\hat{\mathbf{x}}\right)\cdot\mathbf{y}}{\gamma\hbar E_y} \approx -8.5\times10^{-11}\ \left(\mathrm{V/m}\right)^{-1}, \tag{S3.14}$$

which corresponds to

$$\Gamma_0 \approx -6.6\times10^{-12}\ \mathrm{T/(A/m^2)} = -6600\ \mathrm{Oe/(10^{11}\ A/m^2)} = -66\ \mathrm{Oe/(mA/\mu m^2)}. \tag{S3.15}$$

(5) At $E_{\mathrm{F}}$ = –0.84 eV,

$$\frac{\delta \boldsymbol{\mu}^{\mathrm{A}}\left(\mathbf{m}=\hat{\mathbf{x}}\right)\cdot\mathbf{y}}{\gamma\hbar E_y} \approx 3.3\times10^{-10}\ \left(\mathrm{V/m}\right)^{-1}, \tag{S3.16}$$

which corresponds to

$$\Gamma_0 \approx 2.6\times10^{-11}\ \mathrm{T/(A/m^2)} = 26000\ \mathrm{Oe/(10^{11}\ A/m^2)} = 260\ \mathrm{Oe/(mA/\mu m^2)}. \tag{S3.17}$$

Note that these values are much larger than the corresponding effective field strengths for Pt (~50 Oe/$10^{11}$ A/$m^2$) and Ta (~100 Oe/$10^{11}$ A/$m^2$)[3].

Here we remark that the values of $\Gamma_0$, evaluated above based on the DFT, can be revised when the electron correlation is considered. It was reported[18] that the electron correlation rescales the FGT band structure calculated from the DFT by factor $r$ ≈ 1.54; $E_{\mathrm{DFT}}\left(\mathbf{k}\right)/r \approx E_{\mathrm{DMFT}}\left(\mathbf{k}\right)$,

where $E_{\mathrm{DFT}}(\mathbf{k})$ and $E_{\mathrm{DMFT}}(\mathbf{k})$ are the energy dispersions derived from the DFT and the dynamical mean-field theory (DMFT), respectively. The DMFT provides a better account of the electron correlation and $E_{\mathrm{DMFT}}(\mathbf{k})$ agrees well with the energy dispersions of FGT measured by the angle-resolved photoemission spectroscopy[18]. This rescaling affects Figure S13 (and $\Gamma_0$) in two ways; the graphs in Figure S13a,b are squeezed in the horizontal direction by the factor *r* and stretched in the vertical direction by the same factor *r*. The horizontal squeezing is trivial. The vertical stretching arises since the Kubo formula in Eq. (S3.1) contains two powers of energy difference in the denominator and one power of energy in the numerator through the velocity operator $v_j(\mathbf{k}) = (1/\hbar)(\partial H(\mathbf{k})/\partial k_j)$. Moreover, $J_{\mathrm{ex}}$ in the conversion from $\delta\boldsymbol{\mu}^{\mathrm{A}}$ to $\Gamma_0(E_{\mathrm{F}})$ [Eq. (S3.7)] should be scaled down to 1.4/*r*=0.9 eV, which agrees with the exchange interaction strength in the DMFT[18]. Thus $\Gamma_0(E_{\mathrm{F}})$ after the rescaling, which we call $\Gamma_0^{\mathrm{rescale}}(E_{\mathrm{F}})$, is related to the “original” $\Gamma_0(E_{\mathrm{F}})$ by

$$\Gamma_0^{\mathrm{rescale}}(E_{\mathrm{F}}) = \Gamma_0(rE_{\mathrm{F}}). \tag{S3.18}$$

The electron correlation can be correct $\Gamma_0^{\mathrm{rescale}}(E_{\mathrm{F}})$ as well if the correlation modifies the energy eigenstate $|u_{n\mathbf{k}}\rangle$. The energy eigenstate correction is not taken into account in Eq. (S3.18).

To make a direct comparison with our experiment, we need to estimate $E_{\mathrm{F}}$. For an ideal $Fe_3GeTe_2$, $E_{\mathrm{F}}$ is 0 eV. But for real FGT samples, $E_{\mathrm{F}}$ is negative since actual FGT samples tend to be Fe-deficient ($Fe_{3-x}GeTe_2$ with $x > 0$) and the deficiency of positively charged Fe atoms ($Fe^{2+}$ or $Fe^{3+}$) makes FGT hole-doped[19]. For our FGT samples, we estimate that *x* varies between 0.11 and 0.23. If we take the corresponding $E_{\mathrm{F}}$ to be ~–0.2 eV, $\Gamma_0^{\mathrm{rescale}}(E_{\mathrm{F}})$ becomes $\sim 30\ \mathrm{Oe}/(\mathrm{mA}/\mu\mathrm{m}^2)$. Here this number should be regarded as an order of magnitude estimation. Nevertheless, it is encouraging to find that this number we obtained is roughly consistent with the experimentally evaluated value ~50 Oe/(mA/μm$^2$) from the current-induced $H_c$ reduction.

Here we discuss some specific issues related to such SOT in a multilayer FGT. First of all, the situation in our work is very similar to 'the hidden Rashba effect in centrosymmetric systems,' which was previously reported[20]. It was demonstrated that even in centrosymmetric systems, constituent parts of the systems could be subject to the Rasbha effect, although different parts feel the Rashba effect of opposite signs due to the inversion symmetry. A multilayer FGT is center symmetric. Similarly to the hidden Rasbha effect, different layers of a multilayer FGT can be subject to the SOT, although neighboring layers feel the SOT of opposite signs due to the inversion symmetry. In this sense, the SOT in a multilayer FGT can be called hidden SOT. The critical point is that the energy of a multilayer FGT is almost the same regardless of whether neighboring layers are ferromagnetically or antiferromagnetically ordered. It means that the magnetic coupling between adjacent layers is weak, and the dynamics of an individual layer is almost unaffected by the dynamics of its neighboring layers. Thus if the SOT acting on a particular layer can suppress the coercivity of the layer, the coercivity of the multilayer FGT is also suppressed by the SOT, although the SOT alternates in sign between neighboring layers.

The interlayer coupling is usually orders of magnitude weaker than intralayer coupling in many vdW magnets, including FGT[12,21-24]. This weaker interlayer coupling is the direct consequence of the van der Waals interaction across the layer, while the intralayer coupling is predominantly of covalent bonding. This very weak interlayer coupling allows some exciting phenomena to be realized in FGT experimentally, such as the gate-tunable interlayer coupling[23] and the plateau-like magnetoresistance in twisted FGT/FGT junction[24]. Besides, the SOT magnitude has negligible thickness dependence among all our samples (Figure 3e), providing another piece of evidence for a very weak interlayer coupling.

To ensure that the intrinsic SOT is mainly responsible for the measured SOT, one needs to evaluate possible extrinsic SOT effect, although it is challenging to address both experimentally and theoretically. Nonetheless, we remark that there is abundant circumstantial evidence supporting our scenario of the intrinsic SOT. First and foremost, the SOT in ferromagnetic materials is closely related to the anomalous Hall effect, which arises mainly from the intrinsic

Berry curvature, according to a recent study[18]. Moreover, we note a reasonable agreement between the experimentally measured SOT magnitude from the $H_c$ reduction and the calculated intrinsic SOT. If the extrinsic SOT were significantly larger than the intrinsic SOT, the net SOT measured in the $H_c$ reduction experiment should be significantly larger than the calculated intrinsic SOT, which is not the case. Therefore, we argue that intrinsic SOT is the most probable origin of the measured SOT.

Besides, previous work[25] demonstrates the rapid variation of the intrinsic spin Hall conductivity (SHC) of Pt with conductivity, continually changed by incorporating MgO impurities into the Pt. Similarly, we also plot the SHC-conductivity relationship of our FGT samples in Figure S14. The previous study adopted the formula $\sigma_{\mathrm{SH}} = T_{\mathrm{int}}^{-1}\sigma_{\mathrm{SH}}^{*} = T_{\mathrm{int}}^{-1}(\hbar/2e)\mu_0 \mathrm{M_s} t\, \mathrm{H_{DL}}/J * \sigma_{xx}$ to experimentally get the $\sigma_{\mathrm{SH}} - \sigma_{xx}$ relationship for the conventional ferromagnet/Pt SOT system, where the SOT is generated by externally injected spin flow. Differ the case of the previous study, the SOT in FGT arises internally within FGT itself. So we need to make $T_{\mathrm{int}} = 1$ and remove the thickness t, and then get $\sigma_{\mathrm{SH}} \propto \mathrm{H_{SOT}}(\text{per J}) * \sigma_{xx}$ for our FGT system. As shown in Figure S14, the SHC rises about 3-4 times while increasing the conductivity. Such variation is very similar to the behavior for the intrinsic contribution in Pt of the previous study, which provides another piece of evidence for intrinsic SOT in our FGT system. Detailed investigations on the SHC-conductivity relationship with more finely controlled conductivity for FGT would be an exciting research direction in the future.

## Supporting References

## Supporting Figures

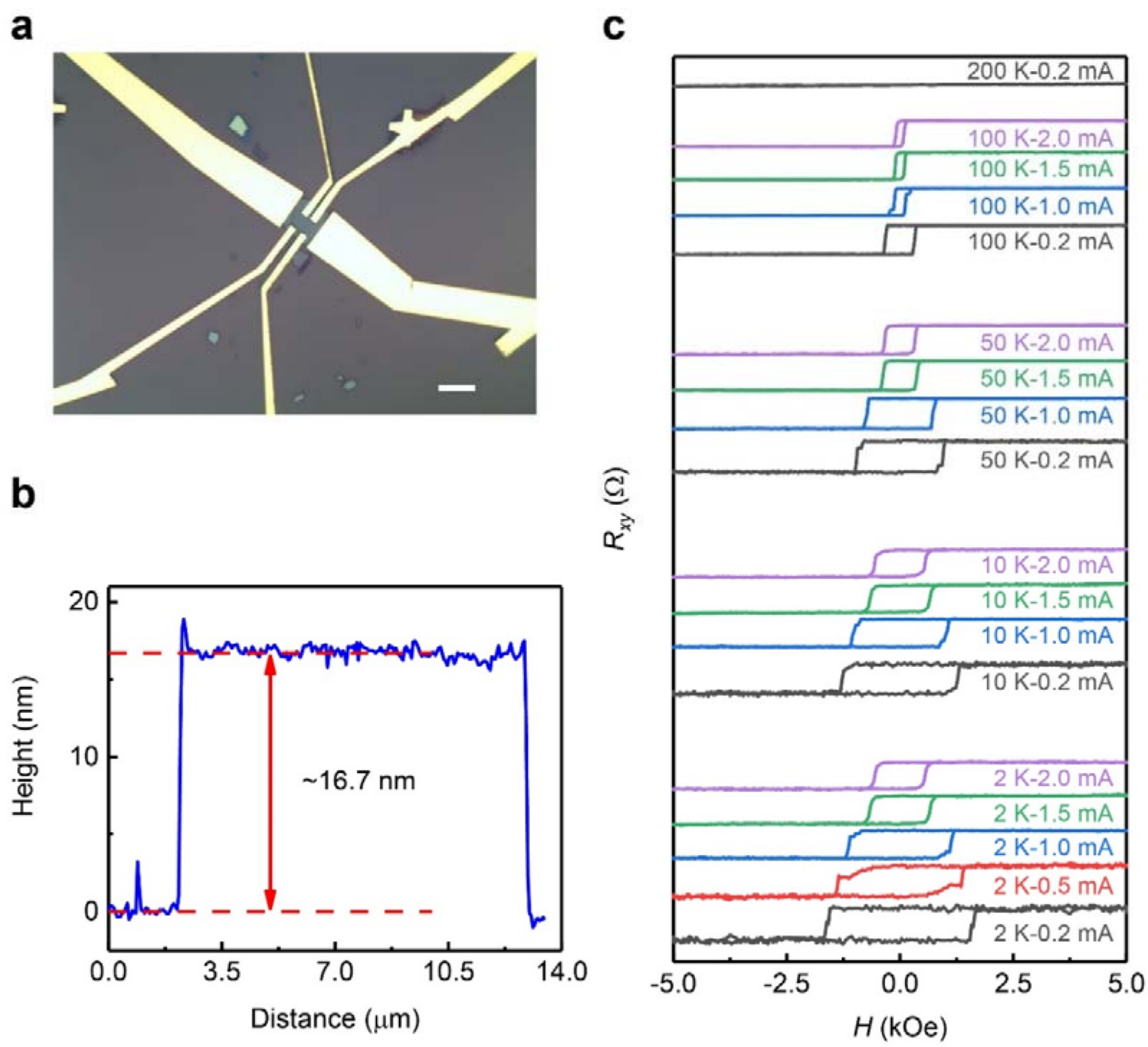

**Figure S1.** Anomalous Hall effect of sample S2. a) Optical image of another FGT nanoflake sample S2 with a Hall-bar geometry electrode. The white scale bar represents 10 μm. b) The thickness of sample S2 is 16.7 nm measured by AFM. c) $R_{xy}$-$H$ curves with various applied current $I$ at 2, 10, 50, 100, and 200 K.

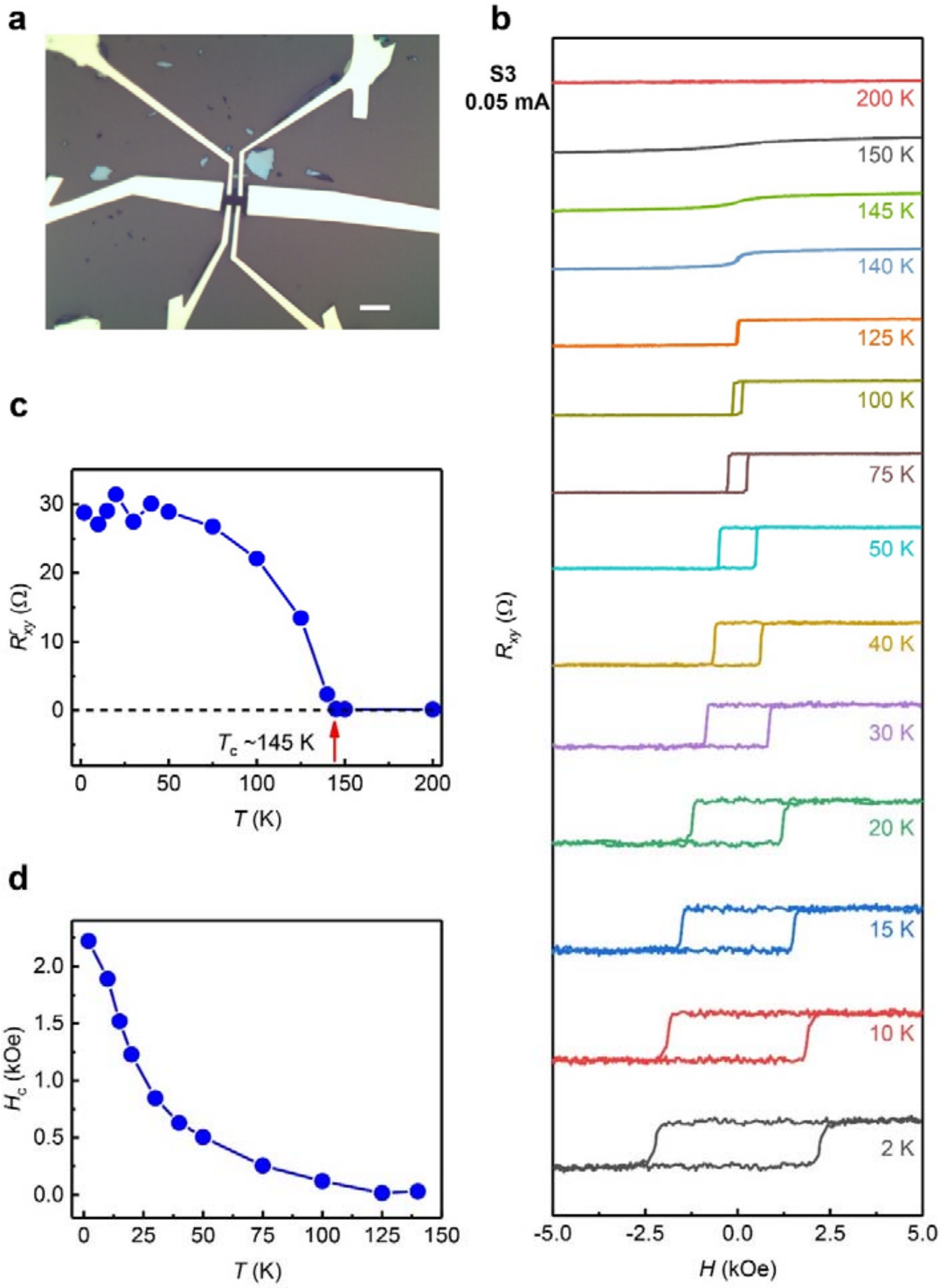

**Figure S2.** Anomalous Hall effect of sample S3 at different temperatures. a) Optical image of another FGT nanoflake sample S3 with a Hall-bar geometry electrode. The white scale bar represents 10 μm. b) $R_{xy}$-$H$ curves with applied current $I$=0.05 mA at various temperatures. c) $R_{xy}^{r}$ as a function of temperature extracted from (b), from which, $T_c$ ~145 K is determined as the onset temperature of $R_{xy}^{r}$. d) $H_c$ as a function of temperature derived from (b).

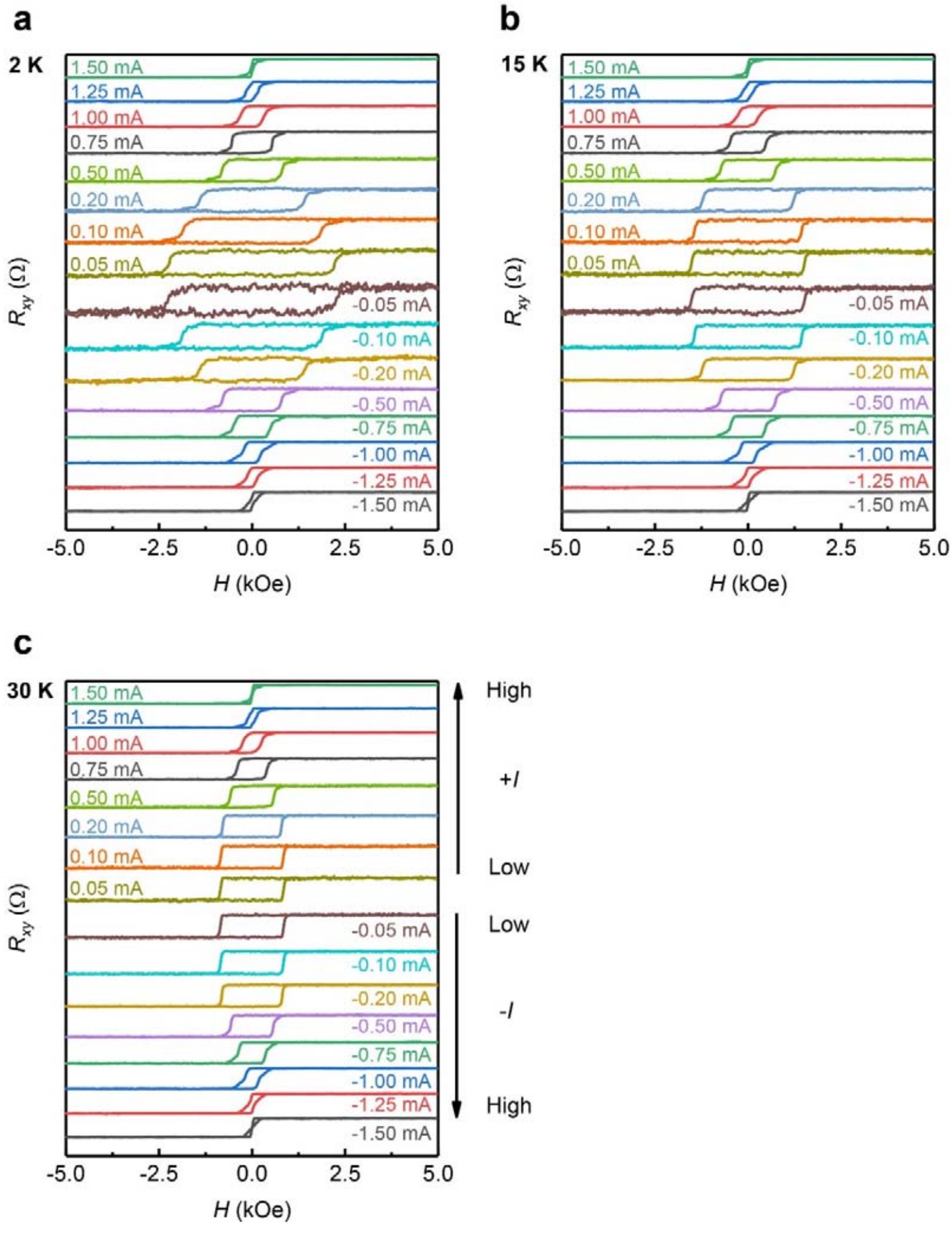

**Figure S3.** Anomalous Hall effect of sample S3 with various currents. a-c) $R_{xy}$-$H$ curves of sample S3 with different applied current $I$ (both $+I$ and $-I$) at 2 (a), 15 (b), 30 K (c), respectively.

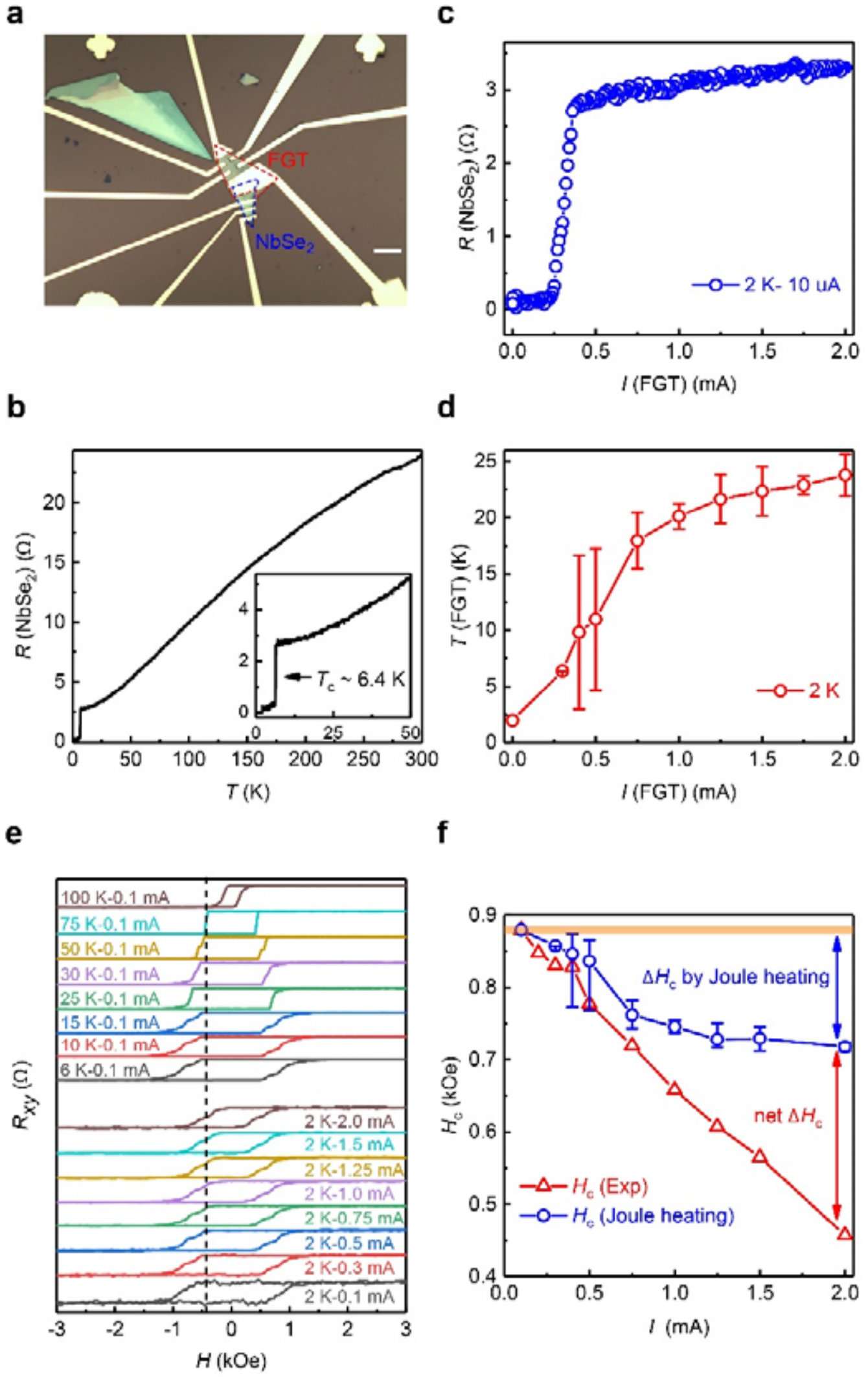

**Figure S4.** Experimental assessment of the Joule heating effect by using $NbSe_2$ as a nanofabricated thermometer with sample S4. a) The optical image of an FGT/$NbSe_2$ heterostructure (sample S4), where $NbSe_2$ is used as a thermometer for FGT. The red and blue dashed boxes indicate the region of FGT and $NbSe_2$, respectively. The white scale bar represents 10 μm. b) $R$($NbSe_2$) as a function of temperature. c) Monitored $R$($NbSe_2$) as a function of applied current on FGT sample, $I$(FGT), at 2 K. d) Calibrated temperature of FGT sample $T$(FGT) by the thermometer at various $I$(FGT) at 2 K. The error bars were calculated from the fluctuation of $R$($NbSe_2$) ~0.1 Ω. e) $R_{xy}$-$H$ curves of sample S4 for various $I$(FGT) at 2 K and multiple temperatures at the fixed small $I$(FGT) = 0.1 mA. f) Experimental $H_c$-$I$ curve (open red triangle) and Joule heating induced $H_c$-$I$ curve (open blue circle). The blue and red arrows indicate the $\Delta H_c$ caused by the Joule heating and the net $\Delta H_c$ without the Joule heating effect, respectively. The error bars in (f) were calculated from the error bars of the calibrated temperature in (d). The large error bars near 0.5 mA in (d) and (f) are due to the almost flat $R$($NbSe_2$)–$T$ relation near 10 K in (b).

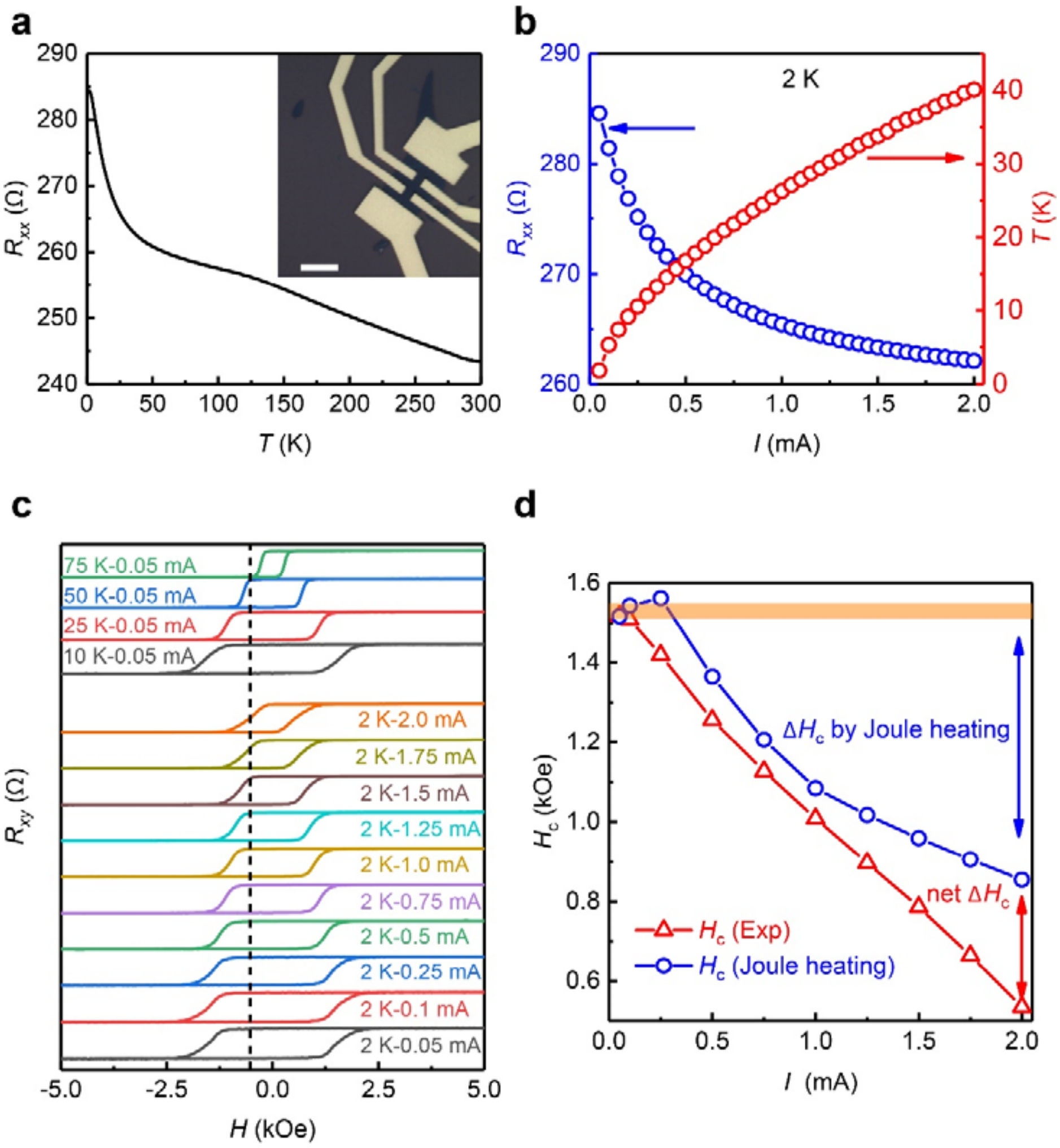

**Figure S5.** Experimental assessment of the Joule heating effect by using $R_{xx}$ as an internal thermometer with sample S5. a) $R_{xx}$ as a function of temperature. The inset is the optical image of the new FGT sample (sample S5), and the white scale bar represents 10 μm. b) Monitored $R_{xx}$ (blue curve) and corresponding calibrated temperature (red curve) of sample S5 as a function of applied current $I$ at 2 K. c) $R_{xy}$-$H$ curves of sample S5 for various $I$ at 2 K and for multiple temperatures at the fixed small $I$ = 0.05 mA. d) Experimental $H_c$-$I$ curve (open red triangle) and Joule heating induced $H_c$-$I$ curve (open blue circle). The blue and red arrows indicate the $\Delta H_c$ caused by the Joule heating and the net $\Delta H_c$ without the Joule heating effect, respectively.

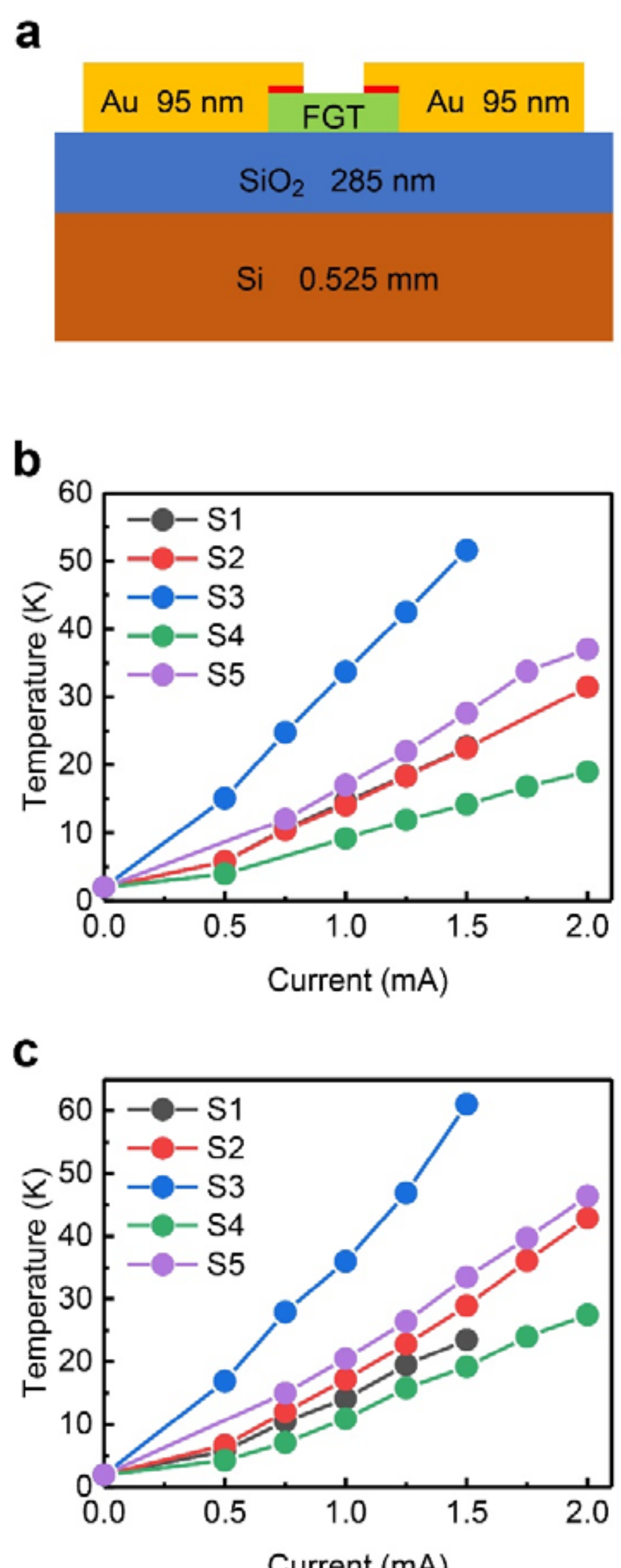

**Figure S6.** COMSOL simulations for Joule heating. a) Schematic of the device geometry used for the COMSOL simulation. Except for the thicknesses specified by the numbers in the figure, other length scales in the figure are only schematic and out of proportion (for instance, FGT of the samples are much thinner than the schematic drawing might imply). b) The calculated temperature of FGT as a function of the applied current for the samples S1, S2, S3, S4, and S5. $\kappa_{OP}$ (out-of-plane thermal conductivity)=$\kappa_{IP}$ (in-plane thermal conductivity) is assumed for FGT in the simulations in (b). c) COMSOL simulations like (b) but with assumption $\kappa_{OP}$=0.01$\kappa_{IP}$. These two sets of calculation results are similar, and the temperature difference between the two sets is within 10 K at 2 mA.

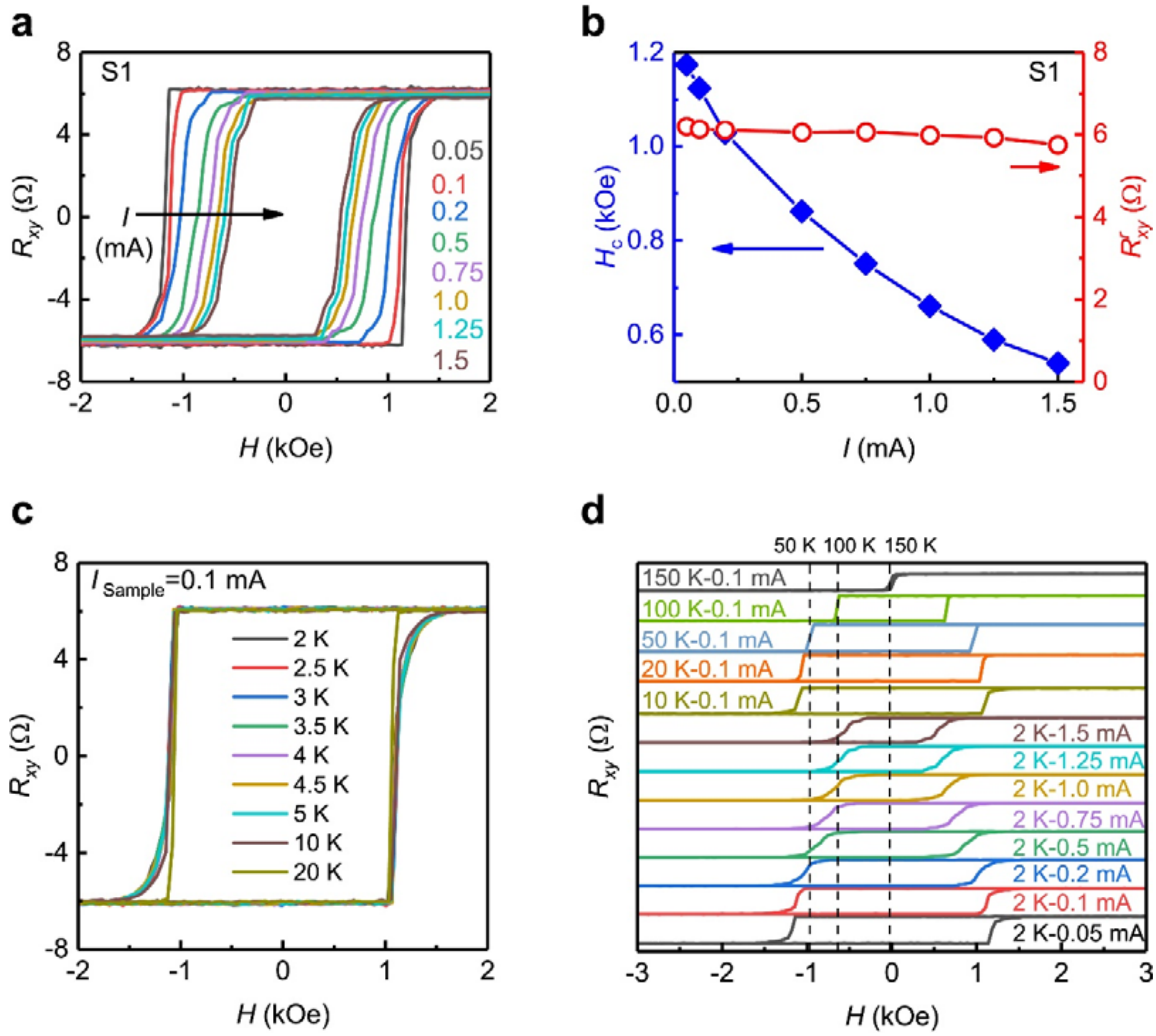

**Figure S7.** Analysis of Joule heating based on the $R_{xy}$-$H$ curves of the sample S1. a) $R_{xy}$-$H$ curves of sample S1 at 2 K with applied current $I$ varying from 0.05 to 1.5 mA. b) Extracted coercive field $H_c$ and remnant Hall resistance $R_{xy}^r$ of sample S1 from (a) as a function of applied current $I$ at 2 K. c) $R_{xy}$-$H$ curves of sample S1 with small $I_{Sample}$= 0.1 mA at various temperatures from 2 to 20 K. d) $R_{xy}$-$H$ curves of sample S1 with multiple $I$ at 2 K and $R_{xy}$-$H$ curves of sample S1 with small $I$=0.1 mA at 10, 20, 50, 100, and 150 K.

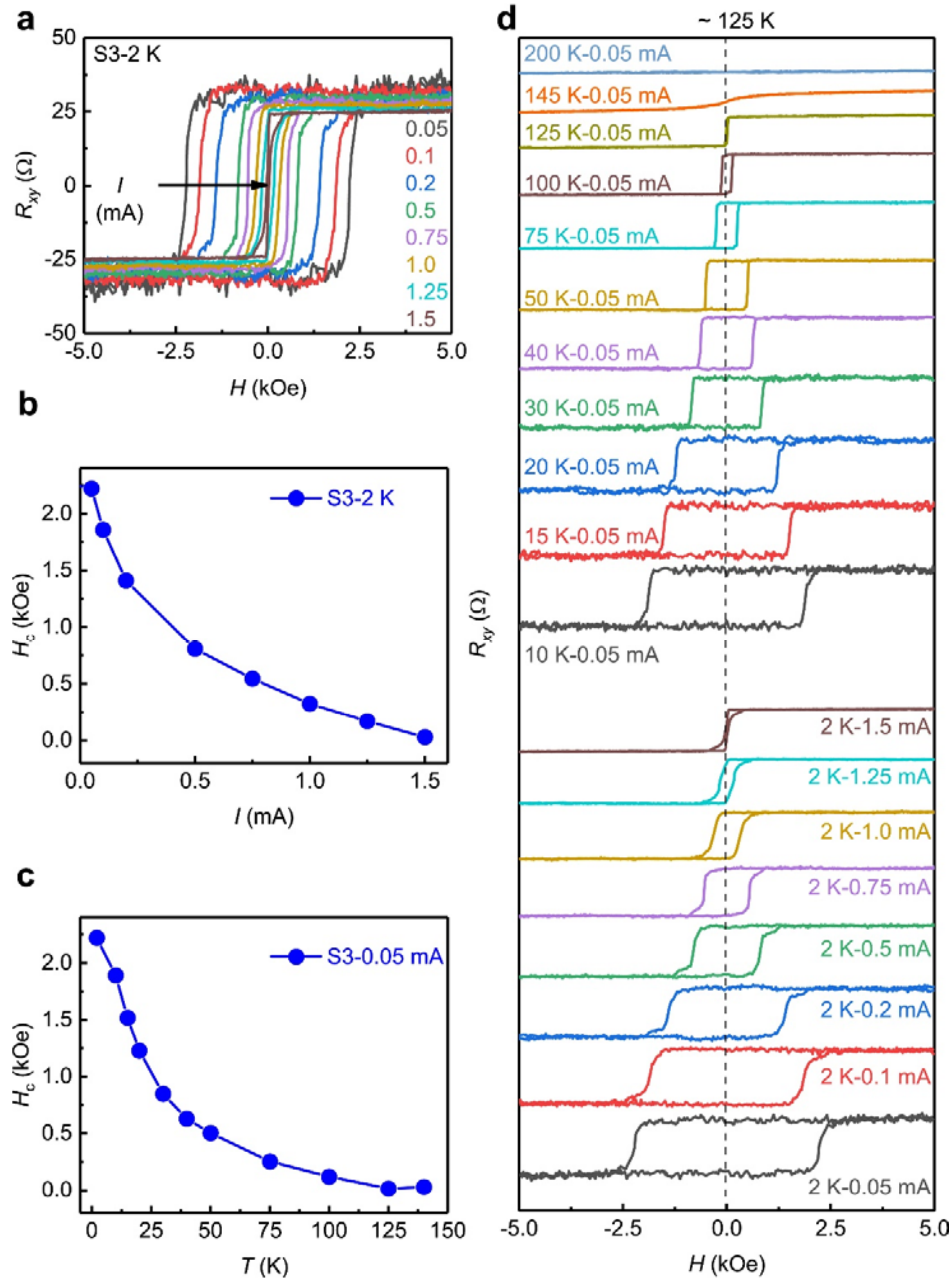

**Figure S8.** Analysis of Joule heating based on the $R_{xy}$-$H$ curves of sample S3. a) $R_{xy}$-$H$ curves of sample S3 at 2 K with applied current $I$ varying from 0.05 to 1.5 mA. b) Extracted coercive field $H_c$ of sample S3 from (a) as a function of applied current $I$ at 2 K. c) Extracted coercive field $H_c$ of sample S3 as a function of temperature with small $I$=0.05 mA. d) $R_{xy}$-$H$ curves of sample S3 with various $I$ at 2 K and $R_{xy}$-$H$ curve of sample S3 with small $I$=0.05 mA at multiple temperatures.

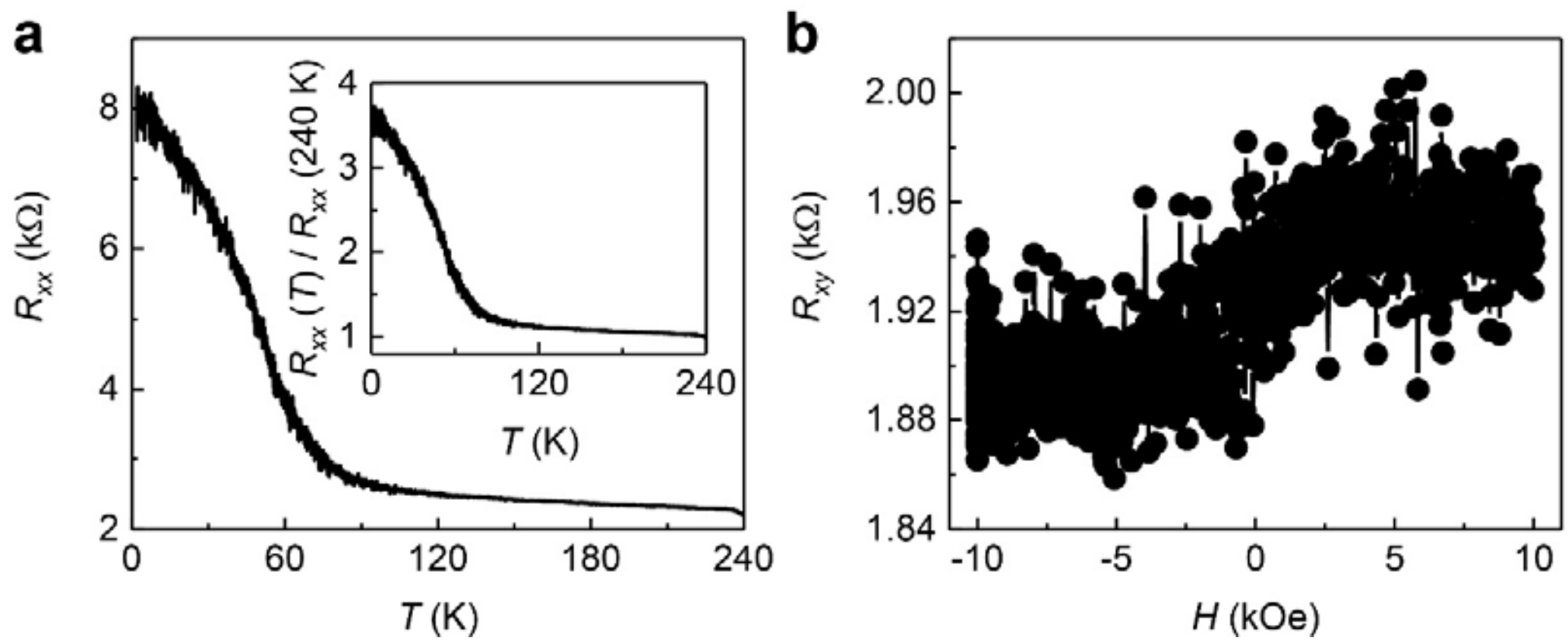

**Figure S9.** The transport properties of a severely damaged FGT sample. a) The resistance $R_{xx}$ as a function of temperature. The inset shows the $R_{xx}(T)/R_{xx}(240\ \mathrm{K}) - T$ curve, where the ratio goes to as high as ~3.7 at 2 K. b) $R_{xy}$-$H$ curve measured at 10 K, where the signal is too noisy to see clear hysteresis loop due to the severe oxidation.

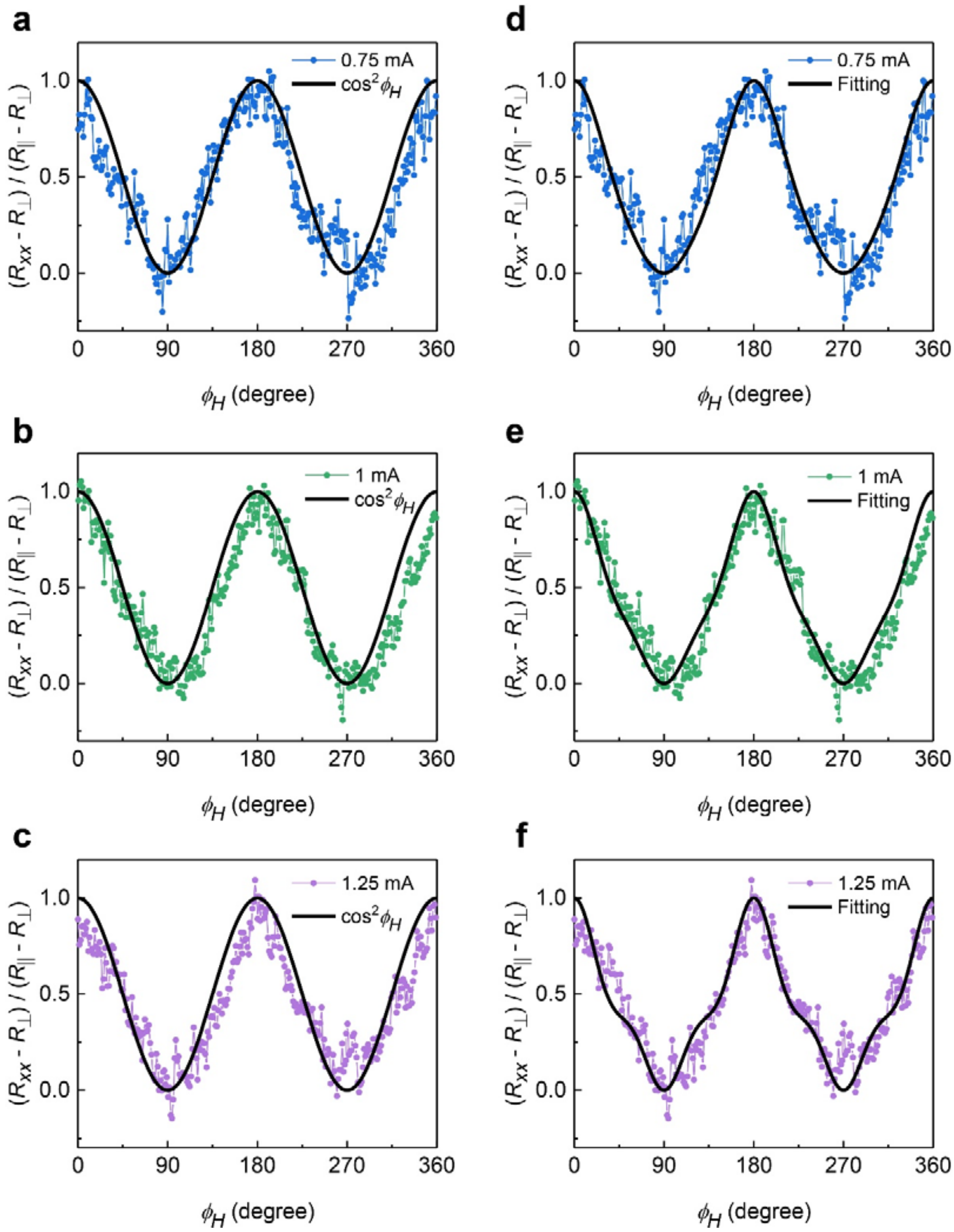

**Figure S10.** AMR measurement. The angular dependence of the normalized longitudinal resistance as a function of $\phi_H$ for the current of 0.75 mA (a,d), 1.0 mA (b,e), 1.25 mA (c,f). The solid black lines in (a,b,c) denote curve of the "conventional" behavior $\cos^2\phi_H$ whereas the black sold lines in (d,e,f) denote the fitting by the multilayer FGT AMR formula (S2-32) in Sec. 2.1.

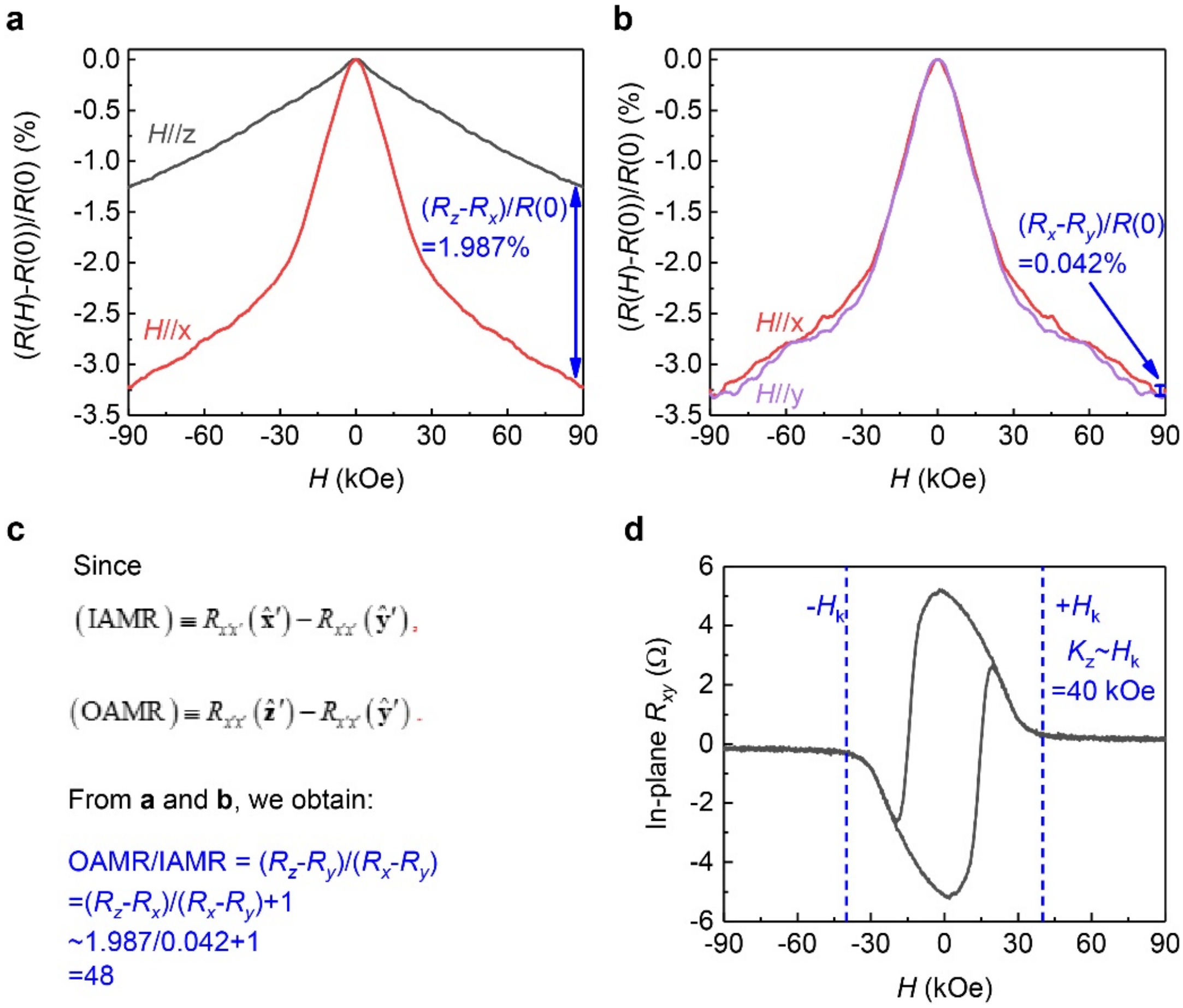

**Figure S11.** The magnetoresistance and in-plane $R_{xy}$ as a function of a magnetic field. a,b) The typical magnetoresistance ratio $(R(H)\text{-}R(0))/R(0)$ as a function of magnetic field $H$ with $H$ applied along the $x$, $y$, $z$ directions at 2 K. The blue arrows indicate the magnetoresistance difference for each direction under 9 T. c) The OAMR/IAMR of FGT is estimated to be ~48 at 2 K and 9 T from (a) and (b), which roughly agrees with our fitted ratio of ~34.2 ± 17.4 by the AMR data, and with the estimated ratio of ~20 at 100 K and 5 T in the previous report[10]. d) The in-plane $R_{xy}$-$H$ curve. We can obtain the in-plane saturated field $H_k$ ~40 kOe. The magnetic anisotropy constant $K_z$ can be roughly estimated to be ~$H_k$, and thus one gets the estimation $K_z$ ~40 kOe. Besides, from comparison with the angle-dependent $R_{xy}$-$H$ curves in the Extended Data Figure 6 of a previous report[12], the misalignment angle $\varphi$ in our AMR experiment should be close to 5˚, which is compatible with the fitted $\varphi$=5.59˚ ± 1.94˚ by the AMR data.

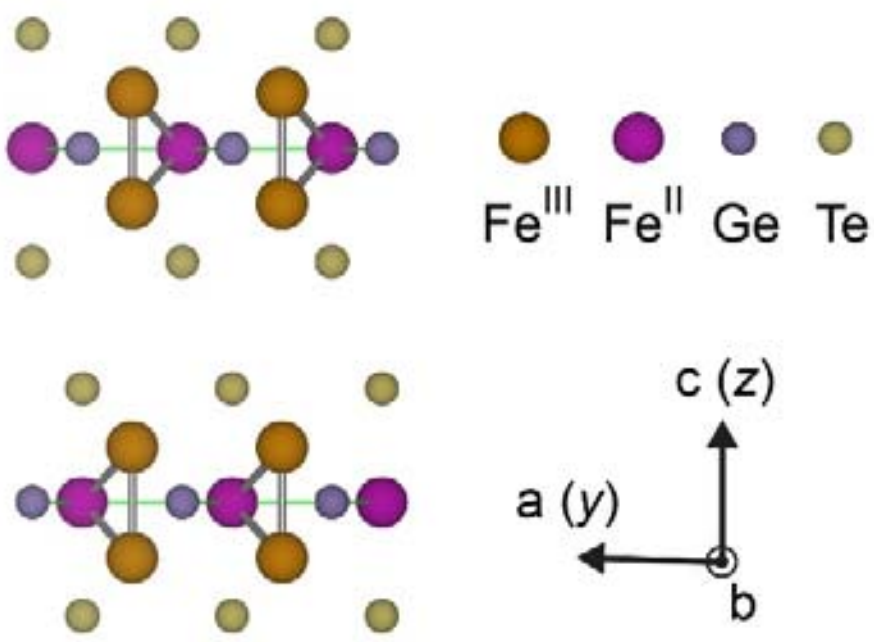

**Figure S12.** Structure of multilayer FGT with AB stacking. Only one A-type layer and one B-type layer are shown for simplicity (viewed along the b axis). These layers repeat as ABABAB... to form an AB-stacked multilayer FGT. Fe$^{III}$ and Fe$^{II}$ denote the two inequivalent Fe sites with oxidation number +3 and +2, respectively.

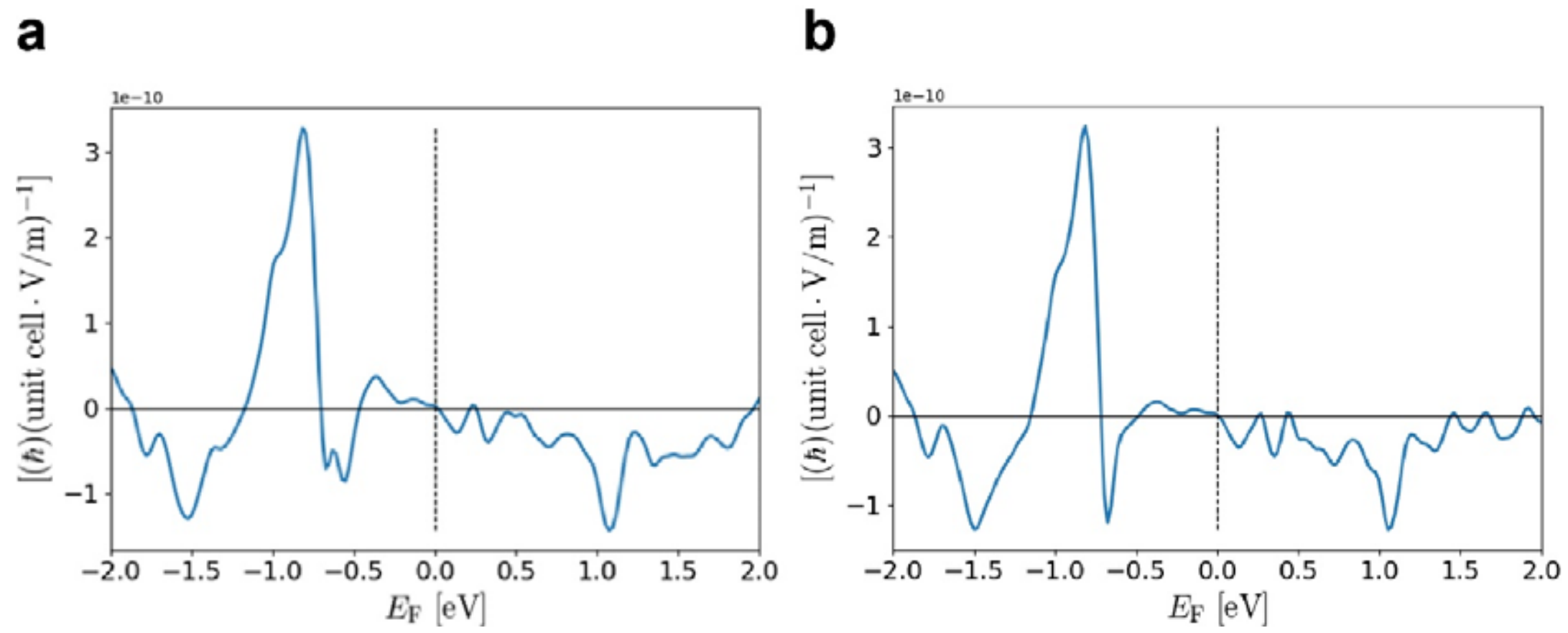

**Figure S13.** Theoretical calculations of the current-induced magnetic moment. a) $\delta\boldsymbol{\mu}^{\text{A}}\cdot\hat{\mathbf{y}}/\gamma E_y$ for $\mathbf{m}=\hat{\mathbf{x}}$, and b) $\delta\boldsymbol{\mu}^{\text{A}}\cdot\hat{\mathbf{x}}/\gamma E_y$ for $\mathbf{m}=\hat{\mathbf{y}}$ evaluated as a function of the Fermi energy $E_\text{F}$.

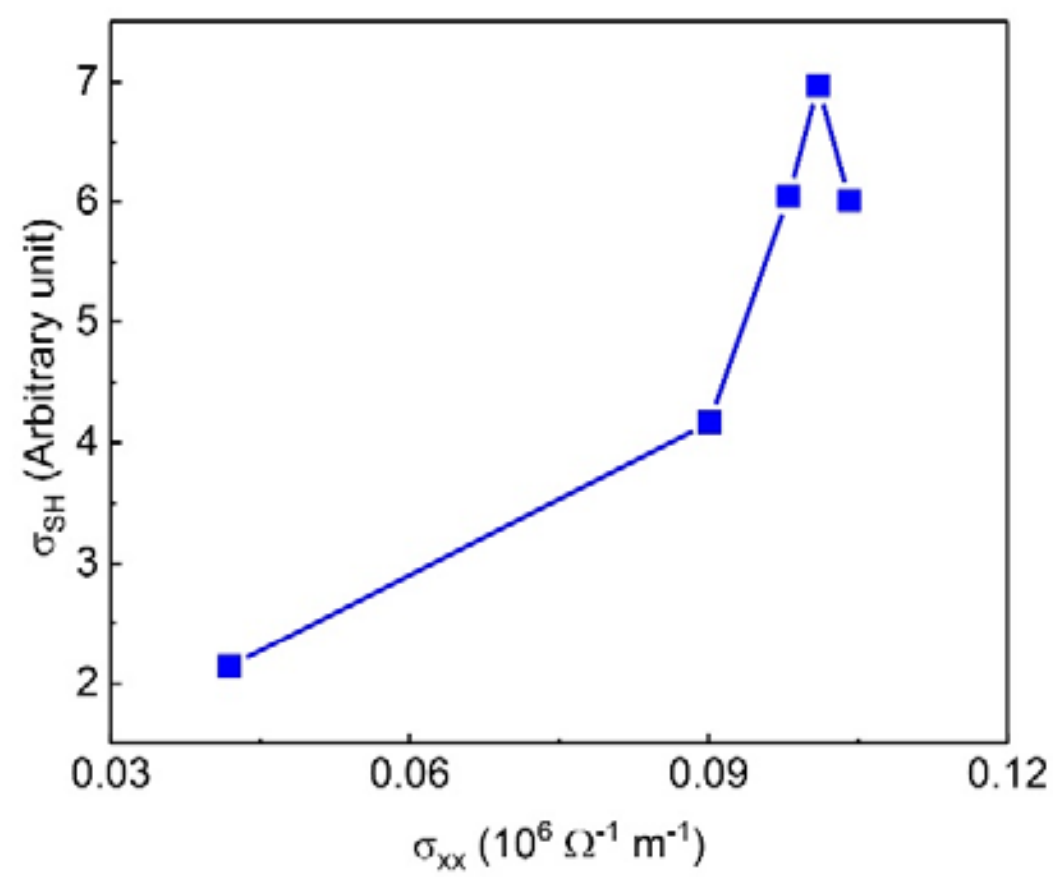

**Figure S14.** Spin Hall conductivity as a function of conductivity for FGT samples.